\documentclass[final,5p,times,twocolumn,authoryear]{elsarticle}

\usepackage{amssymb}
\usepackage{amsmath}
\usepackage{lipsum}
\usepackage{graphicx}
\usepackage{dcolumn}
\usepackage{bm}
\usepackage{hyperref}
\usepackage{float}

\journal{Physics Letters B}

\makeatletter

\newcommand{\Rmnum}[1]{\expandafter\@slowromancap\romannumeral #1@}
\makeatother

\begin{document}

\begin{frontmatter}

\title{Exploring the Cosmological Model Degeneracy with a new evaluate factor $G$}

\author[BNU,IFAA,CO]{Yuan-Bo Xie}
\affiliation[BNU]{
            organization={School of Physics and Astronomy, Beijing Normal University},
            city={Beijing},
            postcode={100875},
            country={China}}
\affiliation[IFAA]{
            organization={Institute for Frontiers in Astronomy and Astrophysics,Beijing Normal University},
            city={Beijing},
            postcode={102206},
            country={China}}
\affiliation[CO]{
            organization={These authors contributed to the work equally and should be regarded as co-first authors.}}
\author[BNU,IFAA,CO]{Yun-Dong Wu}
\author[BNU,IFAA]{Wei Hong}
\author[BNU,IFAA]{Tong-Jie Zhang}
\ead{tjzhang@bnu.edu.cn}

\begin{abstract}
In the context of fitting cosmological models, parameter degeneracy remains a central issue. This paper critically examines traditional methods for constraining parameters and focuses on the $G$ factor as a tool for evaluating the quality of observational data. To ensure analytical independence, two datasets—Cosmic Chronometers (CC) and Baryon Acoustic Oscillations (BAO)—were utilized as samples for parameter fitting, supplemented by Markov Chain Monte Carlo (MCMC) simulations. The Figure of Merit (FoM) matrix served as the final criterion for assessing fitting performance. The results show that the $G$ factor of the CC dataset increases linearly with redshift $z$, whereas the $G$ factor of the BAO dataset follows a cubic relationship. Further analysis indicates that the FoM value for datasets with high $G$ factors is significantly higher than that for datasets with low $G$ factors, thereby validating the $G$ factor's effectiveness as a tool for assessing observational data quality and reducing parameter degeneracy. This suggests that the $G$ factor may serve as a diagnostic tool and selection criterion for optimizing observational datasets in future research.
\end{abstract}


\end{frontmatter}

\section{Introduction}
The theory of General Relativity (GR), proposed by Albert Einstein in 1915 \cite{1915SPAW.......844E}, has driven remarkable progress in addressing a wide range of scientific challenges. Its profound implications have been particularly evident in astronomy, where it successfully explained phenomena such as Mercury’s perihelion precession and the deflection of light during a solar eclipse. Focusing on cosmology—a subfield of astronomy and the central theme of this paper—the theoretical advancements inspired by GR have been transformative. For instance, in 1922, Alexander Friedmann derived the Friedmann-Lemaître-Robertson-Walker (FLRW) metric \cite{1922ZPhy...10..377F}, a solution to Einstein's field equations that provided the mathematical foundation for modern cosmology. This theoretical development was subsequently supported by Edwin Hubble's 1929 observational findings, which established the Hubble Law, demonstrating that the recession velocities of galaxies are approximately proportional to their distances from the Milky Way \cite{1929PNAS...15..168H}. Building on these foundational insights, George Gamow proposed the Big Bang Nucleosynthesis theory in 1946, offering theoretical predictions such as the hydrogen-to-helium abundance ratio and the existence of Cosmic Microwave Background (CMB) radiation \cite{PhysRev.73.803}. These contributions earned him recognition as one of the pivotal figures in modern cosmology.

Over the subsequent decades, cosmology has transitioned into the modern era, marked by the advent of advanced observational tools such as the James Webb Space Telescope \cite{JWST_ERO} and the Planck Satellite \cite{Planck2020}. These instruments have enabled precise testing of theoretical predictions and have deepened our understanding of the universe, particularly by providing high resolution data on the CMB \cite{Planck2020} and observing galaxies at high redshifts \cite{High_z_galaxies}. However, these advancements bring new challenges: What types of data should be prioritized in cosmological research? Furthermore, how should these datasets be curated to maximize reliability, especially given the rising tensions between early and late-universe measurements \cite{HubbletensionSum}? While minimizing observational errors remains a primary objective, the evaluation of cosmological results is inherently multi-dimensional, necessitating the development of standardized methodologies \cite{MCMC_first},\cite{emcee}. \\
This paper addresses these questions by focusing on the specific topic of cosmological parameter constraints, beginning with an overview of the current state of research in this domain. 

Since the 1990s, significant efforts have been made to expand Observational Hubble data (OHD) for constraining cosmological parameters, enabling better simulations of the universe's evolution and insights into its early states. Diverse observational techniques have contributed to this endeavor, including Type Ia supernovae (standard candles) \cite{Riess_1998}, galaxy rotation curves \cite{2003A&A...412..633Z}, and passive galaxy evolution \cite{Marchesini_2009}. These methods have been instrumental in refining the value of the Hubble constant. Moreover, advancements in observational technologies have introduced new dimensions to cosmological research. For example, the COBE satellite provided the first direct observations of the CMB \cite{1992ApJ...396L...1S}, while Baryon Acoustic Oscillations (BAO)—first systematically applied in the Sloan Digital Sky Survey (SDSS) project \cite{2000AJ....120.1579Y}—have emerged as critical tools for parameter studies.

While these methods effectively address the initial question of data selection, there remains a need for a broadly applicable standard to evaluate cosmological models comprehensively. To this end, methodologies addressing model degeneracy and discreteness have been developed, focusing on improving confidence levels in parameter constraints. One central kind of these approaches is the discriminative index \( G \), which serves as a key tool for assessing parameter distinguishability. Building on this concept, \cite{2016RAA....16...50T} refined the definition of \( G \) and introduced the Figure of Merit (FoM) as an additional criterion for model evaluation.

This paper revisits the derivation of the $G$ factor and extends its definition to incorporate curvature density $\Omega_k$, thereby increasing parameter independence and enhancing its FoM. Finally, an effectiveness analysis of the refined $G$ factor will be conducted to determine its suitability as a robust evaluation metric in cosmological parameter studies.

\section{Theoretical foundation}
\subsection{Factor $G$}

This study is based on the $\Lambda$CDM cosmological model. On this foundation, we introduce one of the most critical factors used as an evaluation metric in this research, the $G$ factor, which definition is given by the following formula: 

\begin{equation}
    G_{i} \equiv \frac{1}{\sigma_{\Delta,i}}\frac{\partial H_{th,i}}{\partial \Omega_{\alpha}}, 
\end{equation}

where the index $i$ ranges over the entire set of observational points used in the text, and $\Omega_{\alpha}$ takes three values corresponding to the density parameters $\Omega_{\Lambda},\Omega_{m},\Omega_{k} $, while $\sigma_{\Delta,i} = |H_{obs,i}-H_{th,i}| $ represents the error between the observational values $H_{obs} $ and the theoretical values $H_{th} $ obtained from fitting the $\Lambda$CDM model. Although the observational error of $H$ is generally not equal to the difference between the observed and theoretical values of Hubble parameter, in the context of this study, it is reasonable to assume that the discrepancy between these two errors can be neglected. Therefore, the former is used as a direct substitute for the latter in this case.

From a mathematical perspective, this formula represents the ratio of the partial derivative of the theoretical prediction of the Hubble parameter with respect to the density parameter to its observational error. The partial derivative relationship fundamentally reflects the sensitivity of theoretical predictions derived from physical model to variations in corresponding parameters. Additionally, the division by observational error constitutes a normalization approach, which aims to prevent observation points characterized by high sensitivity but substantial observational errors (and consequently low practical discriminative power) from exerting disproportionate influence on parameter fitting. Finally, statistically, this form is actually the result of logarithmizing the Gaussian likelihood function $L$.
 Such definition has been systematically implemented in \cite{2016RAA....16...50T} and previous research to effectively constrain the confidence intervals of cosmological parameters.
These analytical results suggest that, theoretically, the $G$ factor is well-suited as a statistical measure for assessing the quality of observational points. Therefore, we proceed to further discuss the $G$ factor by explicitly calculating its constituent derivative terms. In the $\Lambda$CDM model, the expression for the Hubble parameter can be written as a function of redshift $z$:

\begin{equation}
    H(z) = H_{0}E(z) = H_{0}\sqrt{\Omega_{\Lambda}+\Omega_{m}(1+z)^{3}+\Omega_{k}(1+z)^2+\Omega_{r}(1+z)^{4}}.
\end{equation}

Here, $H_{0}$ is the Hubble constant, $\Omega_{\Lambda}$ is the cosmological constant density parameter, $ \Omega_{m}$ is the matter density parameter, $\Omega_{r}$ is the radiation density parameter and $\Omega_{k}$ is the curvature density parameter. 
The above four density parameters follow the limitation: $\Omega_{\Lambda}+\Omega_{m}+\Omega_{k}+\Omega_{r} = 1$. Since $\Omega_{r}$ varies extremely little and takes a relatively low value in the late stage of cosmic evolution, the $G$ factor associated with $\Omega_{r}$ is not introduced herein. 
By taking partial derivatives of this expression with respect to each cosmological parameter, we obtain:

\begin{align}
       \frac{\partial H_{th,i}}{\partial \Omega_{\alpha}} 
       & = \frac{H_{0}\Omega_{\alpha}(1+z_{i})^{\beta}}{2 \sqrt{\Omega_{\Lambda}+\Omega_{m}(1+z_{i})^{3}+\Omega_{k}(1+z_{i})^2+\Omega_{r}(1+z_{i})^{4}}} \notag \\
       & = \sigma_{\Delta,i} G_{\alpha,i} ,
       \label{Gk} \\
       \alpha &= \Lambda,m,k, \quad \beta = 0,3,2. \notag
\end{align}

dividing the above results by the observational error at the corresponding observational point yields the $G$ factor for the entire dataset with respect to different parameters.

With the aforementioned results, the question now arises as to how to consolidate these three $G$ factors. In \cite{fishermatrix}, author proposed a method called Fisher information matrix for estimating information density through it, where the matrix element $I(\theta)_{ij}=\partial^{2}lnL/\partial\theta_{i}\partial\theta_{j}$ is close to our definition of the $G$ factor. Then in \cite{jeffreys1939theory},  Jeffreys creatively formulated the prior function in the form of $p(\theta)\propto \sqrt{detI(\theta)}$. Although we do not directly modify the prior here, we can emulate this definition to merge the $G$ factors. Because of the Fisher matrix is diagonalized for independent parameters, a geometric mean definition of the $G$ factor can be expressed as follows:
\begin{equation}
    G = \sqrt[3]{\prod^{\alpha = \Lambda,m,k} {G_{\alpha}}}
\end{equation}
(The cube root operation here is solely intended to ensure the normalization of $G$ factor.)
However, this definition was not employed in the current study due to the fact that, as indicated by Eq.\eqref{Gk}, the contribution of $G_{k}$ becomes overly dominant at higher redshifts, which could compromise the stability of the fitting procedure. To circumvent this issue, we adopt a quadratic mean type definition, specifically:
\begin{equation}
    G = \sqrt{\sum^{\alpha = \Lambda,m,k} {G_{\alpha}^{2}}}.
\end{equation}
This definition essentially performs a physical summation by treating the $G$ factor of each parameter as orthogonal vectors. Such an approach ensures that the $G$ factor maintains a softer variation  even when individual $G$ component exhibit extreme values, while leveraging the information of the $I(\theta)$ (by computing the mean of the matrix trace).
By the latter definition, it allowing us to express the final theoretical form of the $G$ factor as:

\begin{equation}
    G _{th,i} = \frac{H_{0}\sqrt{1+(1+z_{i})^{4}+(1+z_{i})^{6}}}
    {2\bar{\sigma}_{\Delta} \sqrt{\Omega_{\Lambda}+\Omega_{m}(1+z_{i})^{3}+\Omega_{k}(1+z_{i})^2+\Omega_{r}(1+z_{i})^{4}}} .
    \label{G_th}
\end{equation}

It is worth noting that the measurement error for each observational point $\sigma_{\Delta,i}$ is replaced by the overall average error of the data set $\bar{\sigma}_{\Delta}$. This is because $G_{th}$ should serve as a theoretical prediction value, and its predicted curve should be continuous and reflect the properties of the entire data set. Therefore, the measurement error, which reflects the characteristics of individual observation points has been replaced. However, even with the above formula to compute the theoretical prediction, we cannot directly assess the quality of the observation points. Hence, it is necessary to incorporate the observational data into the formula, replacing $H_{0}$ and $\bar{\sigma}_{\Delta}$ with the observational values. In practice, it is nearly impossible to measure extremely low-redshift observation points to directly obtain $H_{0}$. Therefore, we propose a transformation: by multiplying both the numerator and denominator by the expansion rate of the universe $E(z)$, we can further convert these two quantities into forms that can be used with existing observational data for calculation:

\begin{align}
    G _{obs,i} = \frac{H_{0,obs,i}\sqrt{1+(1+z_{i})^{4}+(1+z_{i})^{6}}}
    {2\sigma_{0,obs,i} \sqrt{\Omega_{\Lambda}+\Omega_{m}(1+z_{i})^{3}+\Omega_{k}(1+z_{i})^2+\Omega_{r}(1+z_{i})^{4}}}\\
    = \frac{H_{obs,i}\sqrt{1+(1+z_{i})^{4}+(1+z_{i})^{6}}}
    {2\sigma_{obs,i} \sqrt{\Omega_{\Lambda}+\Omega_{m}(1+z_{i})^{3}+\Omega_{k}(1+z_{i})^2+\Omega_{r}(1+z_{i})^{4}}} .
\end{align}

\subsection{Figure of Merit}
Figure of Merit (FoM) is a quantitative measure of the effectiveness of a system, experiment, method, or theory for a specific task. It is typically used to compare the merits of different methods or systems, or to select the most appropriate solution in design and optimization. The form of FoM can be a simple numerical value or a composite indicator, with its specific definition varying across different studies.

However, FoM, as a criterion in Bayesian inference for data quality, its validity generally requires adherence to the following conditions: 
\\1. Same error distribution: The errors associated with constrained parameters during the inference process must follow the same class of distributions, typically normal distributions. If different distributions are involved, the geometric definition of FoM as an ellipse in the error plane is inherently violated;
\\2. Constraints from likelihood only: The parameters must be mutually constrained exclusively by the relationships specified in the likelihood function. If no constraints exist, the FoM degenerates into a circle, rendering comparative analysis meaningless. Conversely, if additional constraints beyond those in the likelihood function are imposed, the entire constraint framework becomes fundamentally flawed.

In this research, we will adopt a form of FoM previously used in astronomy, which is described as: “The Dark Energy Task Force (DETF) Figure of Merit is the reciprocal of the area of the error ellipse enclosing the 95\% confidence limit in the $\omega_{0}-\omega_{a}$ plane."\cite{2006astro.ph..9591A} In more intuitive mathematical terms, this represents the inverse of the area of the ellipse enclosed by the $2\sigma$ confidence region of the joint Gaussian distribution of two cosmological parameters.
As can be clearly discerned from the definition above, this formulation was designed to facilitate the evaluation of the parameter $\omega$ in the equation of state of Dark Energy. The author intends to establish a ranking method for constraining power of different datasets on $\omega$ through the FoM magnitude. Specifically, datasets demonstrating superior ability in constraining parameters to narrower ranges would attain higher FoM values, thereby guiding subsequent measurement processes based on this criterion. Following this proposition, some studies have employed FoM to assess SNIa datasets \cite{10.1111/j.1365-2966.2008.13380.x}, \cite{J.C.Bueno_Sanchez_2009}. Our study seeks to implement a similar analytical method, albeit with the evaluation target shifted to the OHD dataset \cite{Ma_2011} for the examination to differentiation capability of factor $G$. The specific mathematical definition of the FoM is shown in the appendix section, only the final mathematical form is given here: 

\begin{equation}
    FoM_{XY} = \frac{\pi}{A} = \frac{1}{\sqrt{(1-\rho^{2})}\sigma_{X}\sigma_{Y}}.
    \label{FoM_matrix_element}
\end{equation}

\section{Numerical methods}
\subsection{Data Sources and Processing Methods}
With the above definitions in place, we can now proceed to further discuss the degeneracy of cosmological parameters in the model, starting from the parameter constraints derived from the observational data. It is important to note that this study uses the Cosmic Chronometers (CC) dataset \cite{CC} and the BAO dataset \cite{2021PhRvD.103h3533A}, specific data are presented in Table. \ref{CC data} and Table. \ref{BAO data}. 
The inclusion of these two datasets here stems from their complementary nature, as demonstrated in a prior study \cite{OHD_def}, where the authors systematically consolidated them into a more comprehensive Observational Hubble Data (OHD) compilation for subsequent analyses. In contrast to conventional Standard Candle method, this consolidated dataset incorporates a higher proportion of high redshift observations ($z\sim2$). This enhancement not only ensures improved model stability during parameter constraint procedure, but also amplifies the discrimination ability of the $G$ factor in it (the latter advantage arises from the functional dependence of $G$ on redshift $z$, as explicitly defined in Eq.\eqref{G_th}, where $G \propto (1 + z)^{3/2}$ at high redshifts). 

Since the calculation of the $G$ factor requires the input of multiple parameters dependent on the cosmological model, following the determination of the aforementioned datasets, the next step is to use Monte Carlo Markov chain (MCMC) simulations to infer parameters in the model. The essence of this method lies in combining the Monte Carlo sampling technique with Markov chains, which generates samples from the target distribution to meet the requirements of our simulation. This effectively addresses the drawback of the relatively small sample size of the CC+BAO dataset. A necessary condition for performing MCMC is the input of a likelihood function that reflects the parameter distribution probabilities. Without loss of generality, we can assume that the likelihood function follows a normal distribution:
\begin{equation}
    ln(L_{H}) =\sum_{j=1}^{\{1,2\}}(\sum_{i=1}^{n}{-\frac{[H_{obs,i}-H_{j}(z)]^{2}}{2\sigma_{H_{obs,i}}^{2}} -\frac{1}{2}ln(2\pi\sigma_{H_{obs,i}}^{2} }) ),
\end{equation}
where $ln(L_{H})$ is log-likelihood function, $n$ is the size of whole dataset, $H_{obs}, \sigma_{H_{obs}}$ represent the actual observed values and the standard deviation of the observed values of the Hubble parameter in CC and H(z) represents the theoretical predicted value of the Hubble parameter at the current red-shift. In particular, $H_{1,i}$ and $H_{2,i}$ here refer to $H_{1,i}(z_{i},\theta_{else}|\Omega_{k})$ and $H_{2,i}(z_{i},\theta_{else}|\Omega_{k}=0)$, respectively. 
The employ of double likelihood summation in this context arises from two principal considerations. Firstly, the physical constraint requiring $\Omega_{\Lambda}+\Omega_{m}+\Omega_{k}=1$, which introduces a probability that MCMC sampling of these independent density parameters may fail to generate valid samples satisfying this condition, potentially leading to collapse. The integration of $\Omega_{k}=0$ likelihood component effectively decrease this computational instability. From a theoretical point of view, under the $\Omega_{k}=0$ condition, $\Omega_{k}$ operates on a distinct order of magnitude compared to the other two density parameters $\Omega_{\Lambda}, \Omega_{m}$. This decoupling ensures that their posterior estimations remain essentially unaffected by curvature variations. Consequently, the secondary effects generated from this magnitude disparity to the final curvature determination proves statistically negligible \cite{wu2024cosmicdynamicseinsteincartantheory}.

With the likelihood function in place, multiplying it by a reasonable prior yields the posterior function required for the MCMC sampler. Additionally, it should be noted that the priors in this study include direct constraints on the curvature density parameter. The specific constraint range has been fine-tuned through a series of adjustments, but is initially based on observational data from \cite{Planck2020}. Now presenting the approximate prior range used in the program: $0 < H < 100\ \mathrm{km\ s^{-1}} \mathrm{Mpc^{-1}}$ and $- 0.0012 < \Omega_{k} < 0.0026$. 

\subsection{Feasibility Analysis}
\begin{figure*}[ht!]
    \centering
    \begin{tabular}{ccc}
        \includegraphics[width=0.33\textwidth]{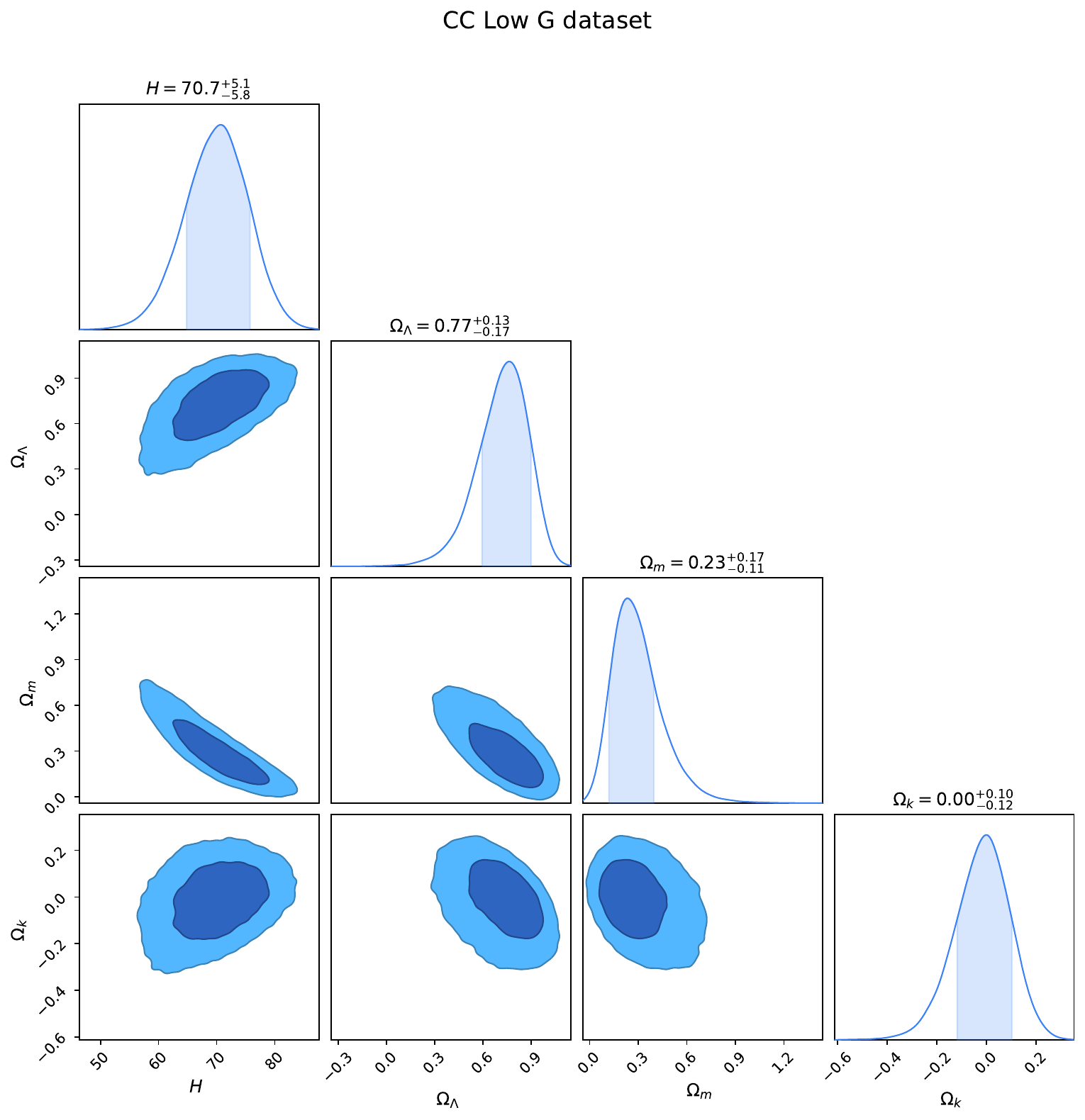} &
        \includegraphics[width=0.33\textwidth]{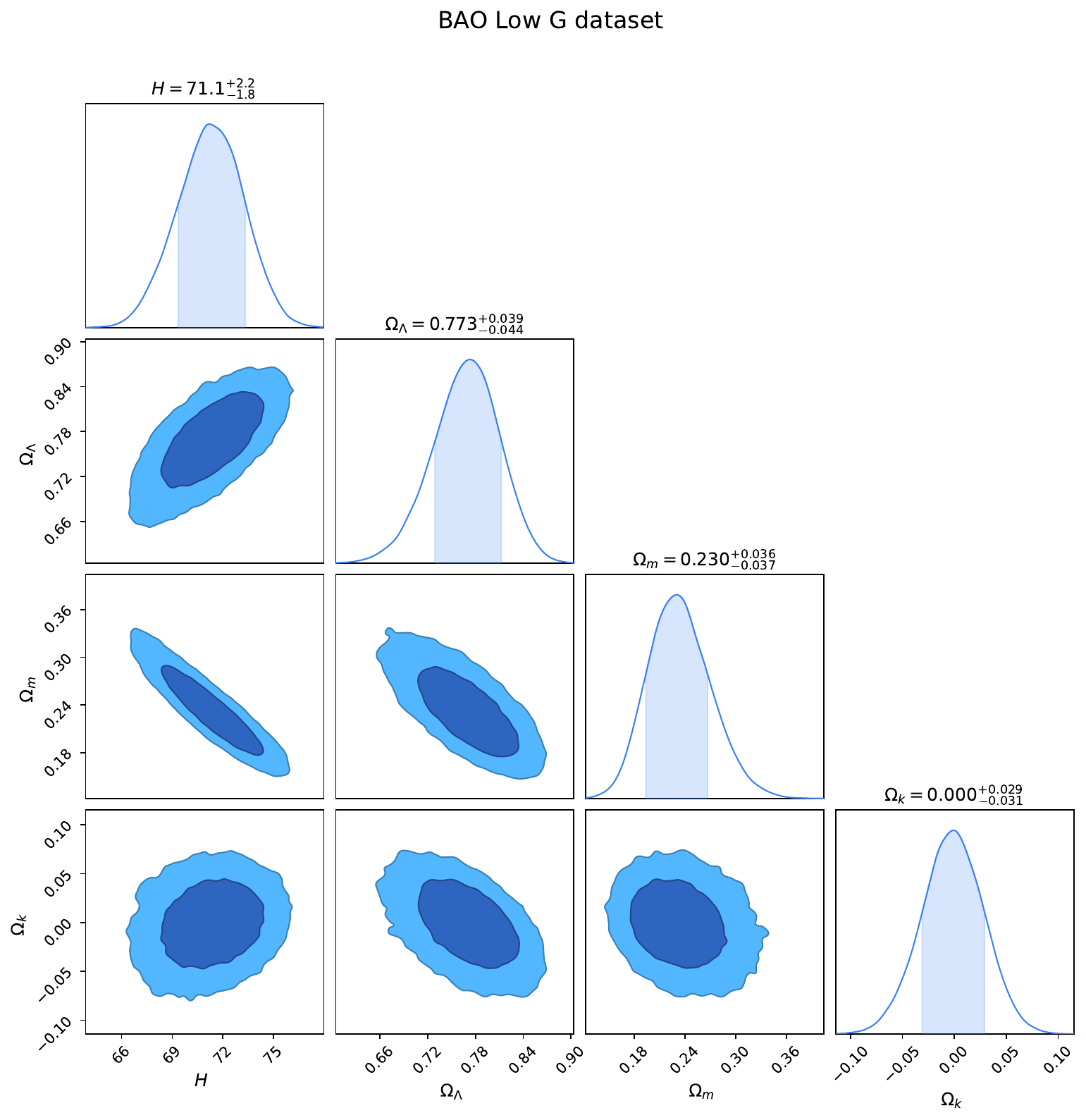} &
        \includegraphics[width=0.33\textwidth]{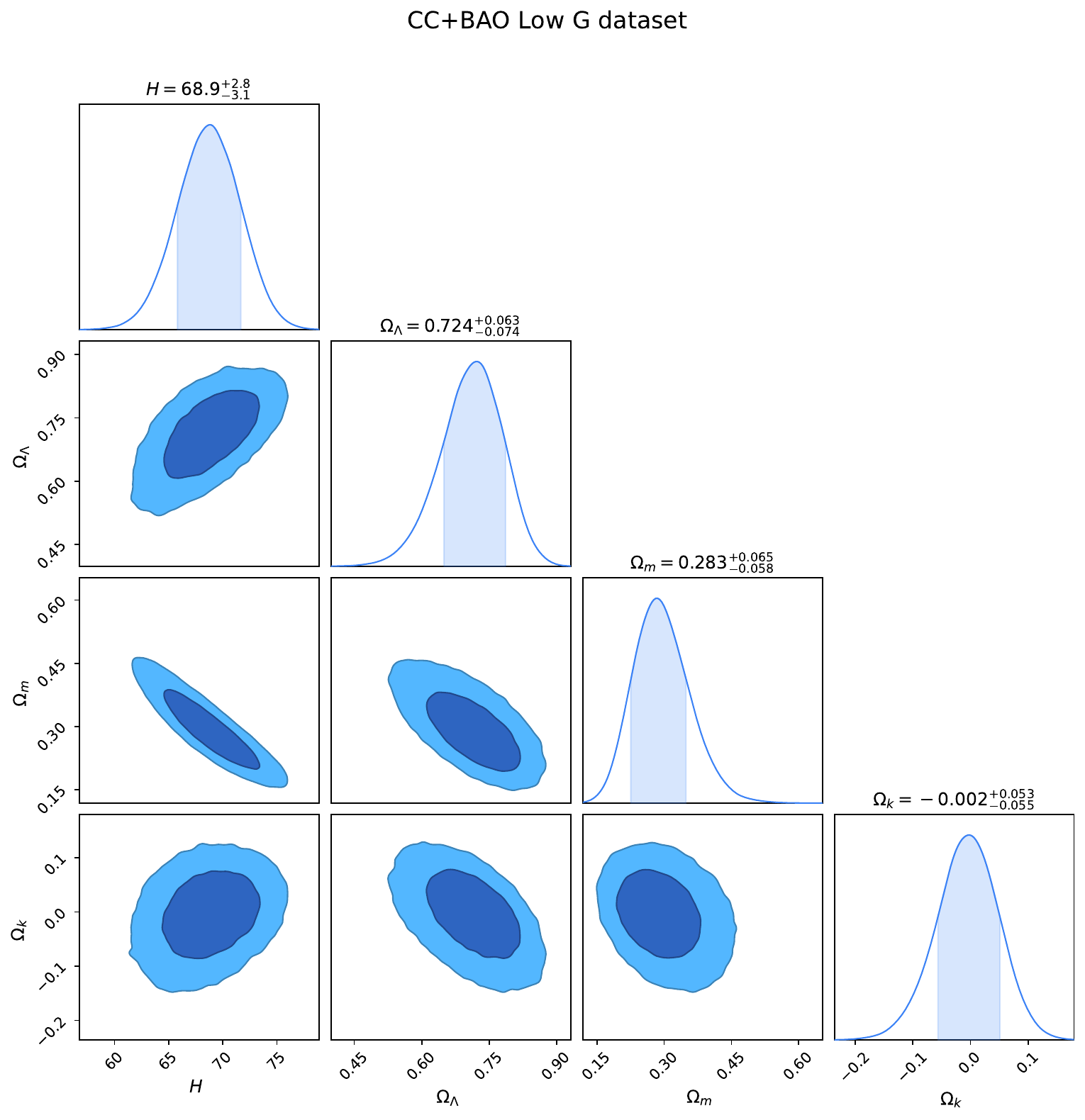} \\
        \includegraphics[width=0.33\textwidth]{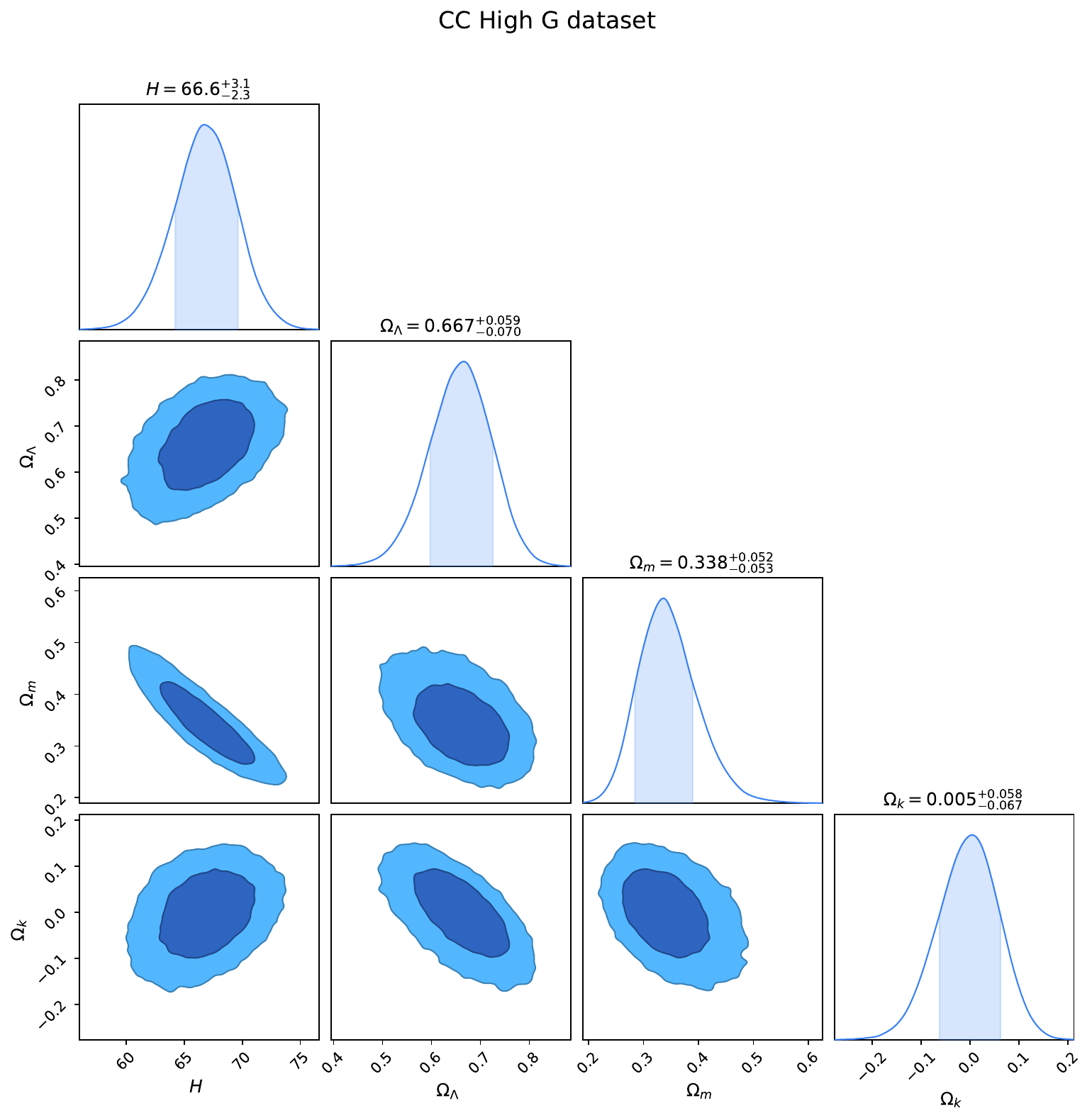} &
        \includegraphics[width=0.33\textwidth]{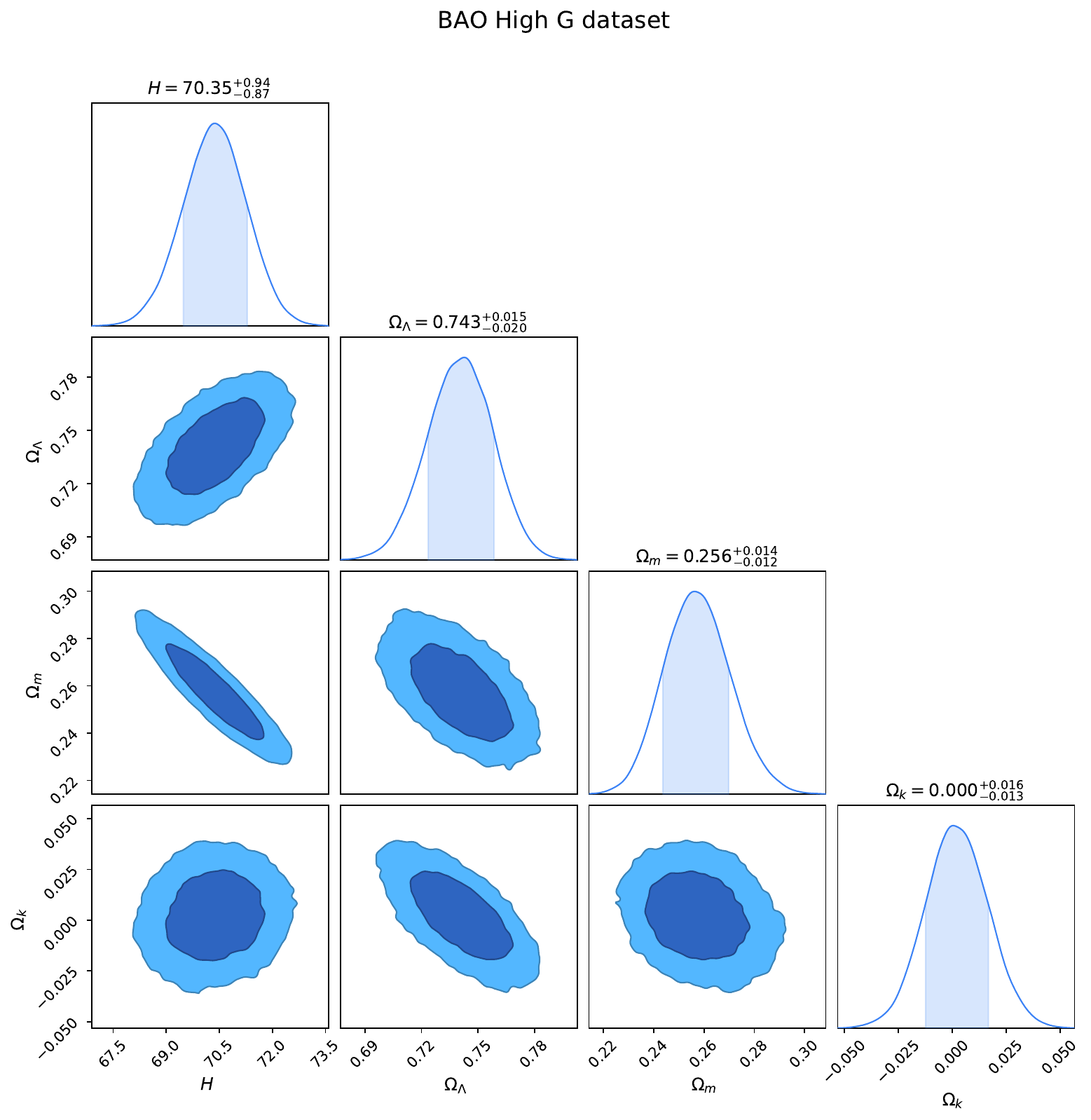} &
        \includegraphics[width=0.33\textwidth]{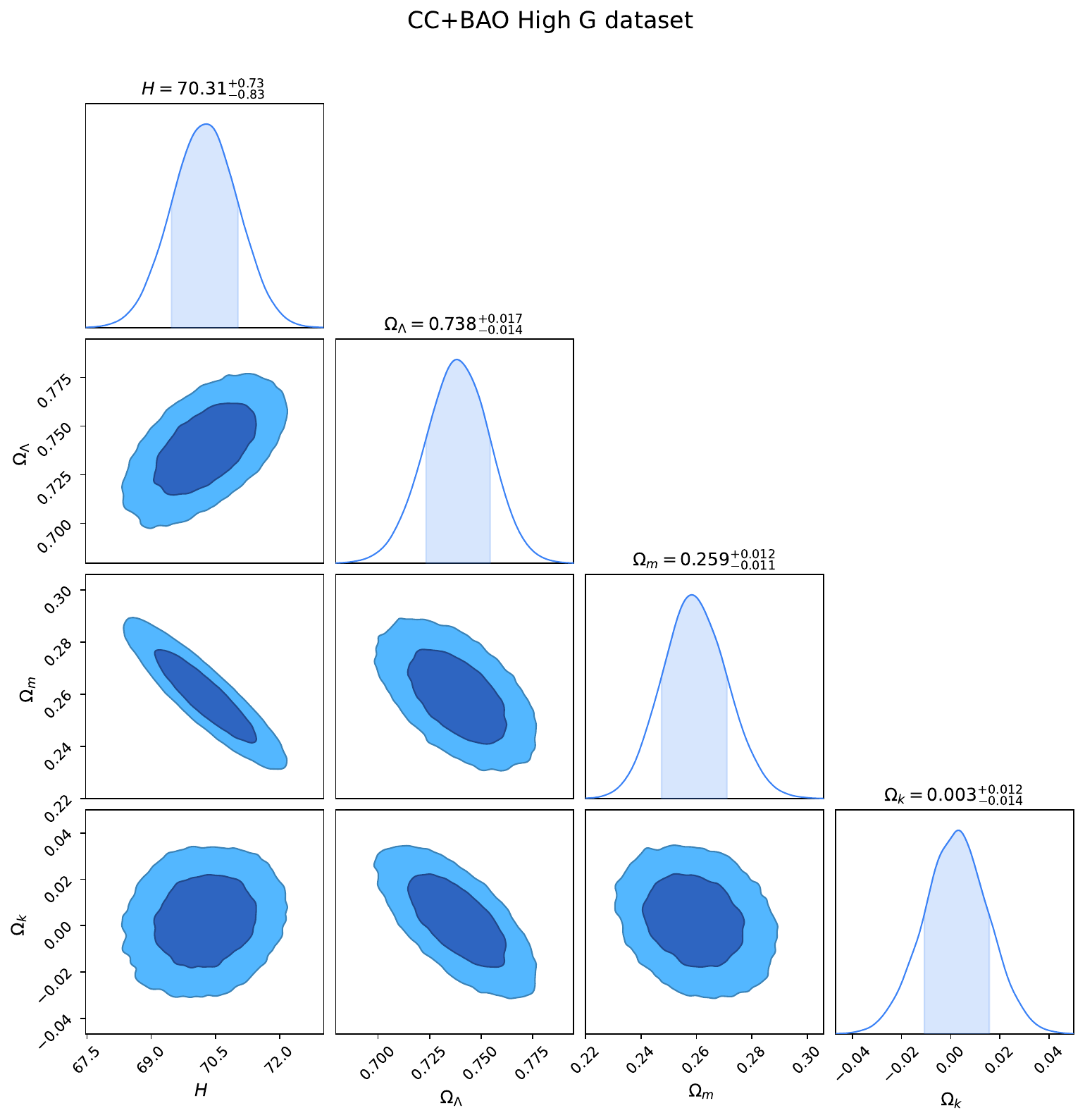}
    \end{tabular}
    \caption{illustrates the posterior distributions from the corresponding MCMC runs in a $3\times2$ matrix layout, representing the data combinations (CC, BAO, CC+BAO) for the two $G$ factor states (Low, High). Within each subplot, the parameters $(H_{0}, \Omega_{\Lambda}, \Omega_{m}, \Omega_{k})$ are uniformly ordered from top to bottom and left to right.}
    \label{fig:simplegrid}
\end{figure*}

In this study, the parameter constraints are implemented using Python, leveraging the MCMC computational capabilities of the Emcee package in \cite{emcee}. The MCMC computations employ 10,000 steps, with an initial burn-in phase of 1,000 steps. The parameter constraint result for all datasets, along with their corresponding Bayesian Information Criterion (BIC) and Reduced Chi-Square ($\chi^{2}_{v}$) evaluations, are detailed in Table.~\ref{BIC}.
\\Initial inspection reveals an anomalous increase in the $\chi^{2}_{v}$ value for the high $G$ dataset compared to the low $G$ dataset. This behavior, however, arises directly from the inherent error-screening effect within the definition of the $G$ factor. Careful examination shows that the parameter uncertainties constrained using the high $G$ dataset are reduced to approximately $\frac{1}{2}$ to $\frac{1}{3}$ of those obtained from the low $G$ dataset. This reduction precisely counterbalances the observed increase in the $\chi_{v}^{2}$ value. We therefore conclude that the inclusion of the $G$ factor imposes no significant influence on the parameter uncertainties derived from the constraint procedure.
\\Regarding the Bayesian Information Criterion (BIC), the first two data pairs (CC, BAO) exhibit $\Delta BIC$ values below the strong-evidence threshold ($\Delta BIC>8$, \cite{Bayes}). This indicates no statistically significant divergence between them. However, a distinct difference emerges in the third pair (CC+BAO): The high $G$ dataset yields a $BIC$ value approximately 26 lower than its low $G$ counterpart. This substantial reduction provides decisive evidence that the $G$ factor enables rigorous data screening. Furthermore, it demonstrates a clear combinatorial advantage in parameter constraints surpassing the cumulative effect expected from either distinct dataset alone.
\\The Gelman-Rubin diagnostic values for the respective MCMC chains are presented in Table.~\ref{GelmanRubin}, all values are below 1.01, confirming convergence to the vicinity of the equilibrium point. Posterior distributions for each parameter set, generated during the MCMC process, are sequentially shown in Figure.~\ref{fig:simplegrid}. These distributions predominantly exhibit sharply peaked symmetric profiles akin to Gaussian distributions. These characteristics satisfy the prerequisite for employing the FoM as a robust validation metric. 
\\Furthermore, all fits within this study are exclusively based on the $\Lambda$CDM cosmological model, which should be understood as referring to the comparative degeneracy among the fitted models derived from different observational results, all under the framework of this cosmological model.

\begin{table}[ht!]
\centering
\caption{Measurements of the Hubble parameter in the table below are derived with the Cosmic Chronometers method in units of $\mathrm{km\ s^{-1}} \mathrm{Mpc^{-1}}$. }

\label{CC data}
\begin{tabular}{cccc}
\hline
\textbf{z} & \textbf{H(z)} & \textbf{$\sigma_{H(z)}$ }  & \textbf{Reference}\\
\hline
0.07  & 69.0  & 19.6   & \cite{2014RAA....14.1221Z}\\
0.09  & 69    & 12     & \cite{2005PhRvD..71l3001S}\\
0.12  & 68.6  & 26.2   & \cite{2014RAA....14.1221Z} \\
0.17  & 83    & 8      & \cite{2005PhRvD..71l3001S}\\
0.179 & 75    & 4      & \cite{2012JCAP...08..006M} \\
0.199 & 75    & 5      & \cite{2012JCAP...08..006M} \\
0.20  & 72.9  & 29.6   & \cite{2014RAA....14.1221Z} \\
0.27  & 77    & 14     & \cite{2005PhRvD..71l3001S} \\
0.28  & 88.8  & 36.6   & \cite{2014RAA....14.1221Z} \\
0.352 & 83    & 14     & \cite{2012JCAP...08..006M} \\
0.38  & 83    & 13.5   & \cite{2016JCAP...05..014M} \\
0.4   & 95    & 17     & \cite{2005PhRvD..71l3001S}\\
0.4004 & 77   & 10.2   & \cite{2016JCAP...05..014M} \\
0.425  & 87.1 & 11.2   & \cite{2016JCAP...05..014M} \\
0.445  & 92.8 & 12.9   & \cite{2016JCAP...05..014M} \\
0.47   & 89.0 & 49.6   & \cite{2017MNRAS.467.3239R} \\
0.4783 & 80.9 & 9      & \cite{2016JCAP...05..014M}\\
0.48   & 97   & 62     & \cite{2010JCAP...02..008S}\\
0.593  & 104   & 13      & \cite{2012JCAP...08..006M}\\
0.68   & 92    & 8       & \cite{2012JCAP...08..006M}\\
0.75   & 98.8  & 33.6    & \cite{2022ApJ...928L...4B}\\
0.75   & 105   & 10.76   & \cite{2023JCAP...11..047J}\\
0.781  & 105   & 12      & \cite{2012JCAP...08..006M}\\
0.8    & 113.1 & 25.22   & \cite{2023ApJS..265...48J}\\
0.875  & 125   & 17      & \cite{2012JCAP...08..006M}\\
0.88   & 90    & 40      & \cite{2010JCAP...02..008S}\\
0.9    & 117   & 23      & \cite{2005PhRvD..71l3001S}\\
1.037  & 154   & 20      & \cite{2012JCAP...08..006M}\\
1.26   & 135   & 65      & \cite{2023AA...679A..96T}\\
1.3    & 168   & 17      & \cite{2005PhRvD..71l3001S}\\
1.363  & 160   & 33.6    & \cite{2015MNRAS.450L..16M}\\
1.43   & 177   & 18      & \cite{2005PhRvD..71l3001S}\\
1.53   & 140   & 14      & \cite{2005PhRvD..71l3001S}\\
1.75   & 202   & 40      & \cite{2005PhRvD..71l3001S}\\
1.965  & 186.5 & 50.4    & \cite{2015MNRAS.450L..16M}\\
\hline
\end{tabular}
\end{table}

\begin{table}[ht!]
\centering
\caption{The measurement of Hubble parameter $H(z)$ in units of $\mathrm{km\ s^{-1}\ Mpc^{-1}}$ with their errors $\sigma_H$ at redshift $z$ obtained from the BAO method.}
\label{BAO data}
\begin{tabular}{cccc}
\hline
\textbf{$z$} & \textbf{$H(z)$} & \textbf{$\sigma_{H(z)}$} & \textbf{Reference}\\ 
\hline
0.24 & 79.69 & 2.99  & \cite{2009MNRAS.399.1663G}\\
0.30 & 81.7  & 6.22  & \cite{2014MNRAS.439.2515O}\\
0.31 & 78.17 & 4.74  & \cite{2017MNRAS.469.3762W}\\
0.34 & 83.8  & 3.66  & \cite{2009MNRAS.399.1663G}\\
0.35 & 82.7  & 8.4   & \cite{2013MNRAS.435..255C}\\
0.36 & 79.93 & 3.39  & \cite{2017MNRAS.469.3762W}\\
0.38 & 81.5  & 1.9   & \cite{2017MNRAS.470.2617A}\\
0.40 & 82.04 & 2.03  & \cite{2017MNRAS.469.3762W}\\
0.43 & 86.45 & 3.68  & \cite{2009MNRAS.399.1663G} \\
0.44 & 82.6  & 7.8   & \cite{2012MNRAS.425..405B}\\
0.44 & 84.81 & 1.83  & \cite{2017MNRAS.469.3762W} \\
0.48 & 87.79 & 2.03  & \cite{2017MNRAS.469.3762W} \\
0.51 & 90.4  & 1.9   & \cite{2017MNRAS.470.2617A} \\
0.52 & 94.35 & 2.65  & \cite{2017MNRAS.469.3762W} \\
0.56 & 93.33 & 2.32  & \cite{2017MNRAS.469.3762W}\\
0.57 & 87.6  & 7.8   & \cite{2013MNRAS.433.3559C} \\ 
0.57 & 96.8  & 3.4   & \cite{2014MNRAS.441...24A} \\ 
0.59 & 98.48 & 3.19  & \cite{2017MNRAS.469.3762W} \\ 
0.60 & 87.9  & 6.1   & \cite{2012MNRAS.425..405B} \\ 
0.61 & 97.3  & 2.1   & \cite{2017MNRAS.470.2617A} \\
0.64 & 98.82 & 2.99  & \cite{2017MNRAS.469.3762W} \\ 
0.73 & 97.3  & 7.0   & \cite{2017MNRAS.469.3762W} \\ 
0.978 & 113.72 & 14.63 & \cite{2019MNRAS.482.3497Z} \\ 
1.23  & 131.44 & 12.42 & \cite{2019MNRAS.482.3497Z} \\ 
1.526 & 148.11 & 12.71 & \cite{2019MNRAS.482.3497Z} \\ 
1.944 & 172.63 & 14.79 & \cite{2019MNRAS.482.3497Z} \\ 
2.3   & 224    & 8    & \cite{2013AA552A96B} \\ 
2.33  & 224    & 8    & \cite{2017AA603A12B} \\ 
2.34  & 222    & 7    & \cite{2015AA...574A..59D} \\ 
2.36  & 226    & 8    & \cite{2014AA...563A..54P} \\ 
\hline
\end{tabular}
\end{table}

\begin{table*}[ht!]
\centering
\caption{sums the results of parameter constraints obtained via MCMC method for all datasets, along with the corresponding goodness-of-fit metrics: the reduced $\chi^{2}$ and the Bayesian Information Criterion (BIC). The Hubble constant $H_{0}$ is expressed in units of $\mathrm{km\ s^{-1}\ Mpc^{-1}}$. Critically, the dataset size $n$ is determined through the following method: First, compute the $G$ factor for each entry in the input dataset and sort all entries in ascending order of $G$. If the resulting ordered dataset contains an odd number of entries, remove the entry with the smallest $G$ value. The remaining entries are then partitioned equally into two subgroups from midway: the low $G$ subgroup and the high $G$ subgroup. Consequently, the subgroup size $n$ always equals $[N_{CC,BAO,CC+BAO}/2]$ to ensure a balanced sample size.}
\label{BIC}
\begin{tabular}{ccccccc}
\hline
\textbf{Dataset} & \textbf{CC low $G$ } & \textbf{CC high $G$} & \textbf{BAO low $G$} & \textbf{BAO high $G$} & \textbf{CC+BAO low $G$} & \textbf{CC+BAO high $G$}\\ 
\hline
$n$ & 17 & 17 & 15 & 15 & 32 & 32 \\
$\chi^{2}_{v}$ & 0.732 & 1.815 & 0.111 & 0.220 & 0.588 & 1.141 \\
$BIC$ & 151.86 & 157.00 & 103.23 & 102.25 & 263.86 & 237.96 \\
$H_{0}$ & $70.7^{+5.1}_{-5.8}$ & $66.6^{+3.1}_{-2.3}$ & $71.1^{+2.2}_{-1.8}$ & $70.35^{+0.94}_{-0.87}$ & $68.9^{+2.9}_{-3.1}$ & $70.31^{+0.73}_{-0.83}$ \\
$\Omega_{\Lambda}$ & $0.77^{+0.13}_{-0.17}$ & $0.667^{+0.059}_{-0.070}$ & $0.773^{+0.039}_{-0.044}$ & $0.743^{+0.015}_{-0.020}$ & $0.724^{+0.063}_{-0.074}$ & $0.738^{+0.017}_{-0.014}$ \\
$\Omega_{m}$ & $0.23^{+0.17}_{-0.11}$ & $0.338^{+0.052}_{-0.053}$ & $0.230^{+0.036}_{-0.037}$ & $0.256^{+0.014}_{-0.012}$ & $0.283^{+0.065}_{-0.058}$ & $0.259^{+0.012}_{-0.011}$ \\
$\Omega_{k}$ & $0.00^{+0.10}_{-0.12}$ & $0.005^{+0.058}_{-0.067}$ & $0.000^{+0.029}_{-0.031}$ & $0.000^{+0.016}_{-0.013}$ & $-0.002^{+0.053}_{-0.055}$ & $0.003^{+0.012}_{-0.014}$ \\
\hline
\end{tabular}
\end{table*}

\begin{table*}[ht!]
\centering
\caption{presents the Gelman-Rubin diagnostic results for all MCMC processes in this study, which evaluate whether the chains have reached equilibrium \cite{GelmanRubin}. Specific calculation methods and implementation details follow \cite{arviz}.}
\label{GelmanRubin}
\begin{tabular}{ccccccc}
\hline
\textbf{GR Diagnostic} & \textbf{CC low $G$ } & \textbf{CC high $G$} & \textbf{BAO low $G$} & \textbf{BAO high $G$} & \textbf{CC+BAO low $G$} & \textbf{CC+BAO high $G$}\\  
\hline
$\hat{R}_{H_{0}}$ & 1.007 & 1.006 & 1.006 & 1.008 & 1.004 & 1.006 \\
$\hat{R}_{\Omega_{\Lambda}}$ & 1.006 & 1.005 & 1.004 & 1.004 & 1.005 & 1.007 \\
$\hat{R}_{\Omega_{m}}$ & 1.008 & 1.005 & 1.007 & 1.009 & 1.003 & 1.005 \\
$\hat{R}_{\Omega_{k}}$ & 1.006 & 1.007 & 1.007 & 1.006 & 1.005 & 1.007 \\
\hline
\end{tabular}
\end{table*}

\section{Data Analyze}
   \begin{figure*}[ht!]
   \centering
   \includegraphics[width=0.85\textwidth, angle=0]{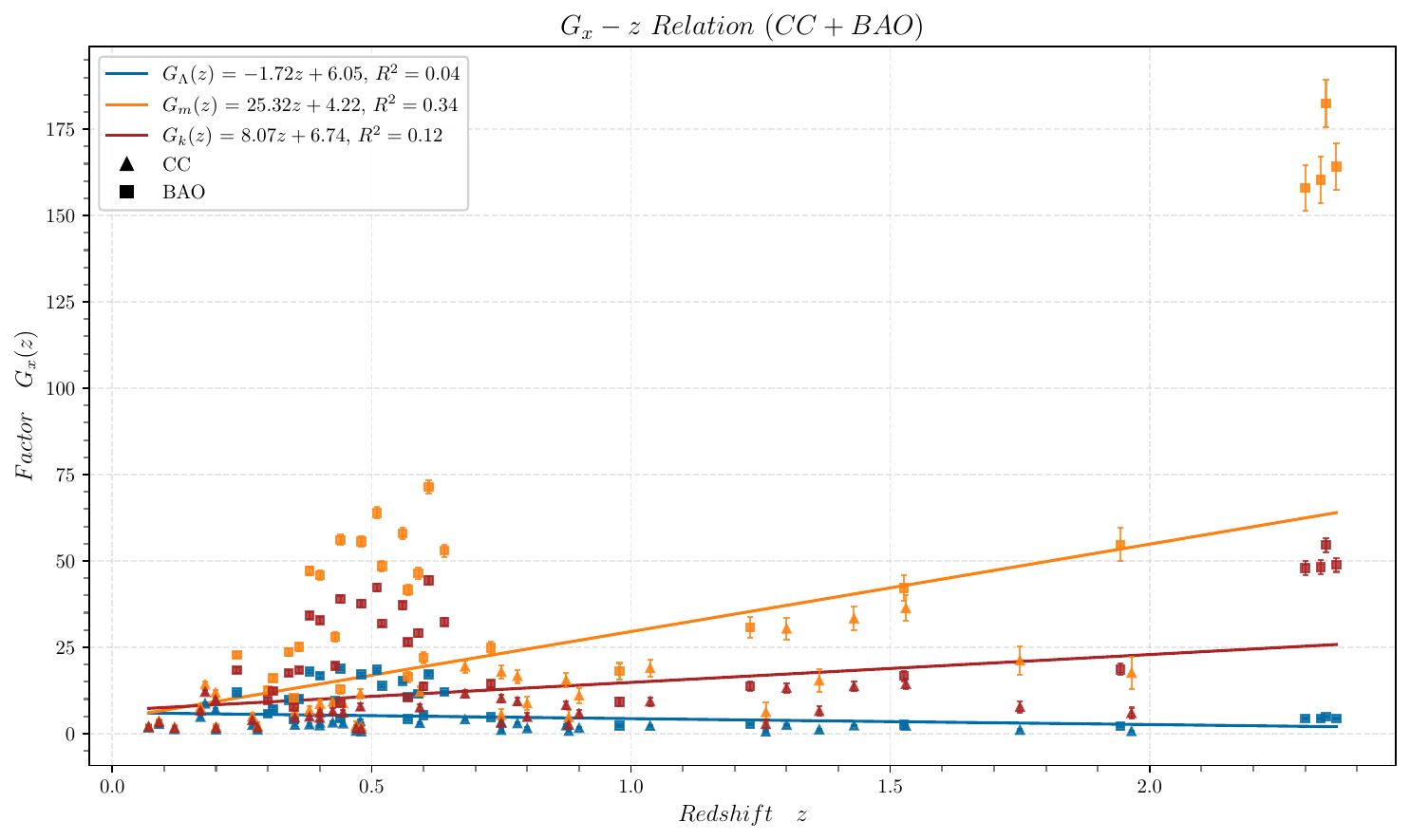}
   \caption{presents scatter plots with error bars for the three components of the \( G \) factor (\( G_{\Lambda} \), \( G_{m} \), and \( G_{k} \)) as functions of redshift \( z \), derived from the CC+BAO dataset. The corresponding linear fitting curves for each data group and regression coefficient $R$ are also shown in the legend and are distinguished by three different colors.}
   \label{Fig1}
   \end{figure*}
\subsection{Fitting the \( G \) factor in CC \& BAO dataset}
In this section, we present and analyze the visualization results for the CC and BAO datasets, obtained using the evaluation metrics and computational methods detailed in the preceding sections.

However, we clarify here that the purpose of fitting curves to the $G$ factor is to establish its evolutionary pattern with redshift $z$. These curves could optimize future observations, targeting the lowest feasible redshifts to minimize resource requirements, while simultaneously achieving improved data quality (represented by higher $G$ values in this study).

Fig.~\ref{Fig1} depicts the linear fits for the three $G$ factor components, $\Omega_{\Lambda}$, $\Omega_{m}$, and $\Omega_{k}$, computed using the combined CC and BAO datasets. As shown, the regression coefficients for all three fits do not achieve statistical significance. A distinct clustering of data points is observed around redshift $z=0.5$. Further analysis reveals that this concentration is primarily comprised of BAO data. Consequently, we conclude that this feature cannot be adequately explained by a linear model. Therefore, a more appropriate approach is to analyze the BAO and CC datasets separately, employing distinct polynomial models for each.

   \begin{figure*}[ht!]
   \centering
   \includegraphics[width=0.85\textwidth, angle=0]{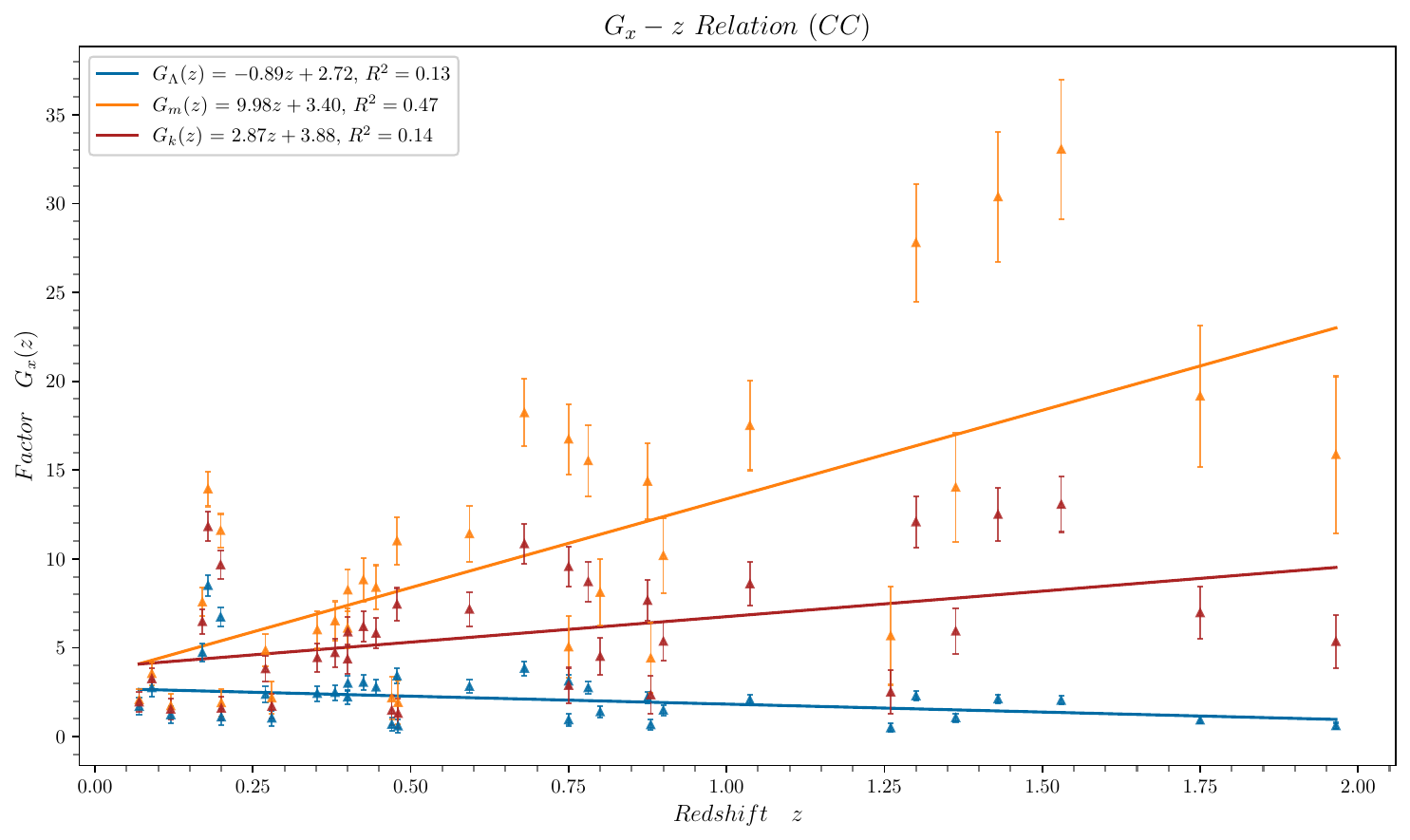}
   \caption{is similar to Fig.~\ref{Fig1} but shows only the scatter plots with error bars for the three components of the \( G \) factor (\( G_{\Lambda} \), \( G_{m} \), and \( G_{k} \)) as functions of redshift \( z \), derived solely from the CC dataset. The legend follows the same format as in Fig.~\ref{Fig1}.}
   \label{Fig2}
   \end{figure*}

   \begin{figure*}[ht!]
   \centering
   \includegraphics[width=0.85\textwidth, angle=0]{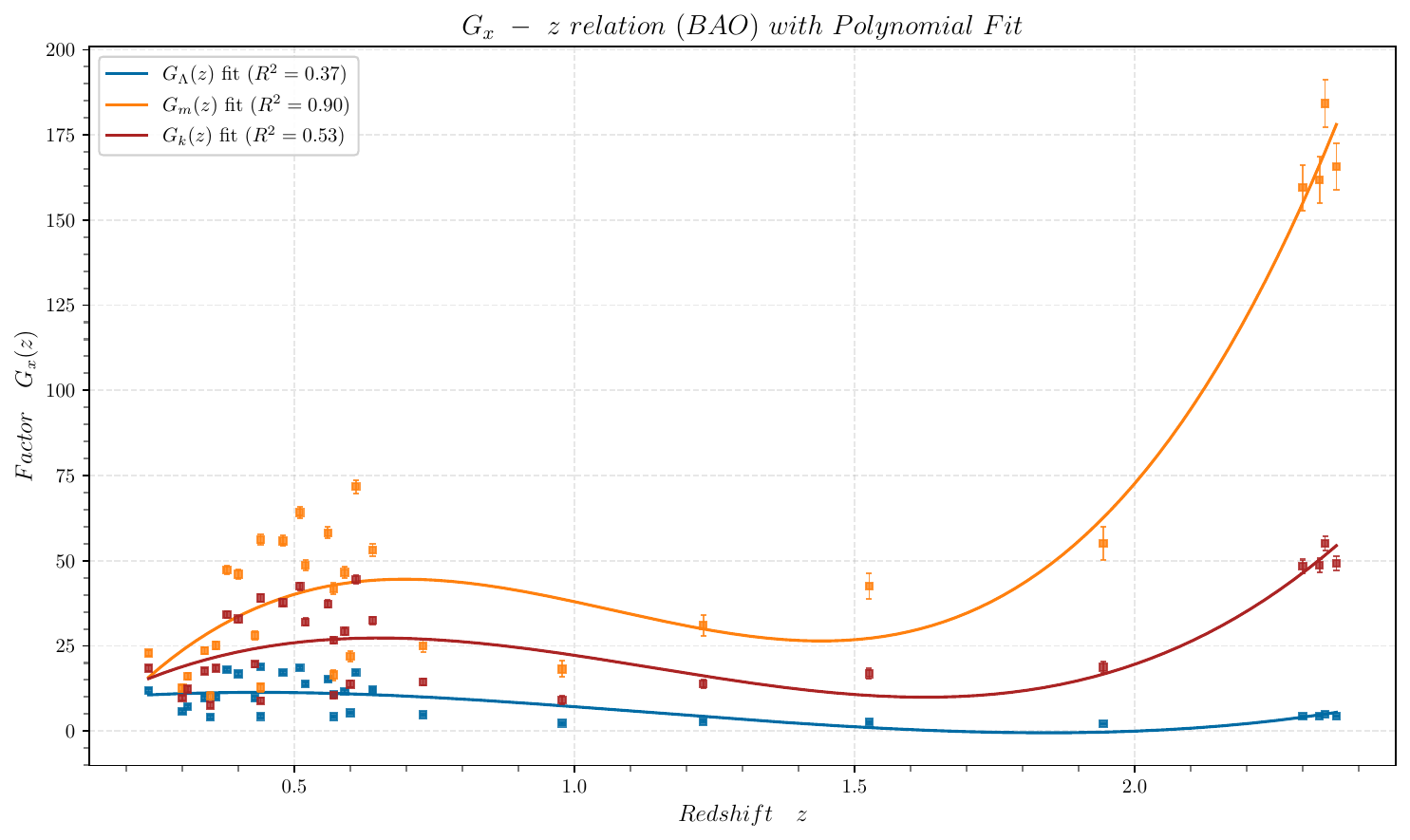}
   \caption{differs from Fig.~\ref{Fig1} in that the dataset is changed to BAO data, and the linear fitting is replaced with a cubic function model, and this adjustment is applied to all components of the \( G \) factor (\( G_{\Lambda} \), \( G_{m} \), and \( G_{k} \)), leading to a notable improvement in fitting accuracy compared to Fig.~\ref{Fig1}. The legend format remains the same as in Fig.~\ref{Fig1}.}
   \label{Fig3}
   \end{figure*}

The adoption of polynomial fitting here is intrinsically related to the definition of the $G$ factor. Since $G$ fundamentally comprises partial derivatives of the Hubble parameter $H$, it follows mathematically that $G$ is linear with the first-order term in the Taylor expansion of $H$. This naturally motivates the extension to higher-order expansions, corresponding to the polynomial forms employed in subsequent fitting. Moreover, we specifically require the fitted curve to exhibit at least one redshift interval demonstrating an inverse correlation between redshift $z$ variation and $G$ factor variation. Such a feature enables $G$ to provide actionable guidance for observational strategies (e.g., avoiding observations in declining regimes) and facilitates validation through simulated data. Polynomial fitting satisfies these requirements while maintaining calculation simplicity.

Fig.~\ref{Fig2} and Fig.\ref{Fig3} present the results of linear fitting using only the CC dataset and polynomial fitting using only the BAO dataset, respectively. While the regression coefficient $R$ (Fig.~\ref{Fig2}) shows only marginal improvement compared to Fig.~\ref{Fig1}, the anomalous clustering observed in the combined dataset (Fig.~\ref{Fig1}) is effectively eliminated. In Fig.~\ref{Fig3}, the regression coefficient $R$ for the $G_{m}$ component reaches a statistically significant strong correlation ($R^{2} > 0.85$), demonstrating the efficacy of the polynomial fitting approach. A maximum polynomial degree of 3 was employed, as simulations indicated that higher-order fits yielded negligible improvements in $R$. Therefore, in accordance with Occam's razor, higher-order terms were excluded.

The figures above reveal a clear ordering of the $G$ factor magnitudes: $G_m > G_k > G_{\Lambda}$. The relatively smaller influence of $G_k$ on model degeneracy compared to $G_m$ is expected, given the tight constraints already placed on curvature. However, the near-horizontal fitting curve for $G_{\Lambda}$ warrants further investigation. Physically, this behavior may reflect a relatively weak dependence of the cosmological model on the dark energy component, as evidenced by the high degeneracy observed in any fit involving $G_{\Lambda}$.

\begin{figure*}[ht!]
   \centering
   \includegraphics[width=0.85\textwidth, angle=0]{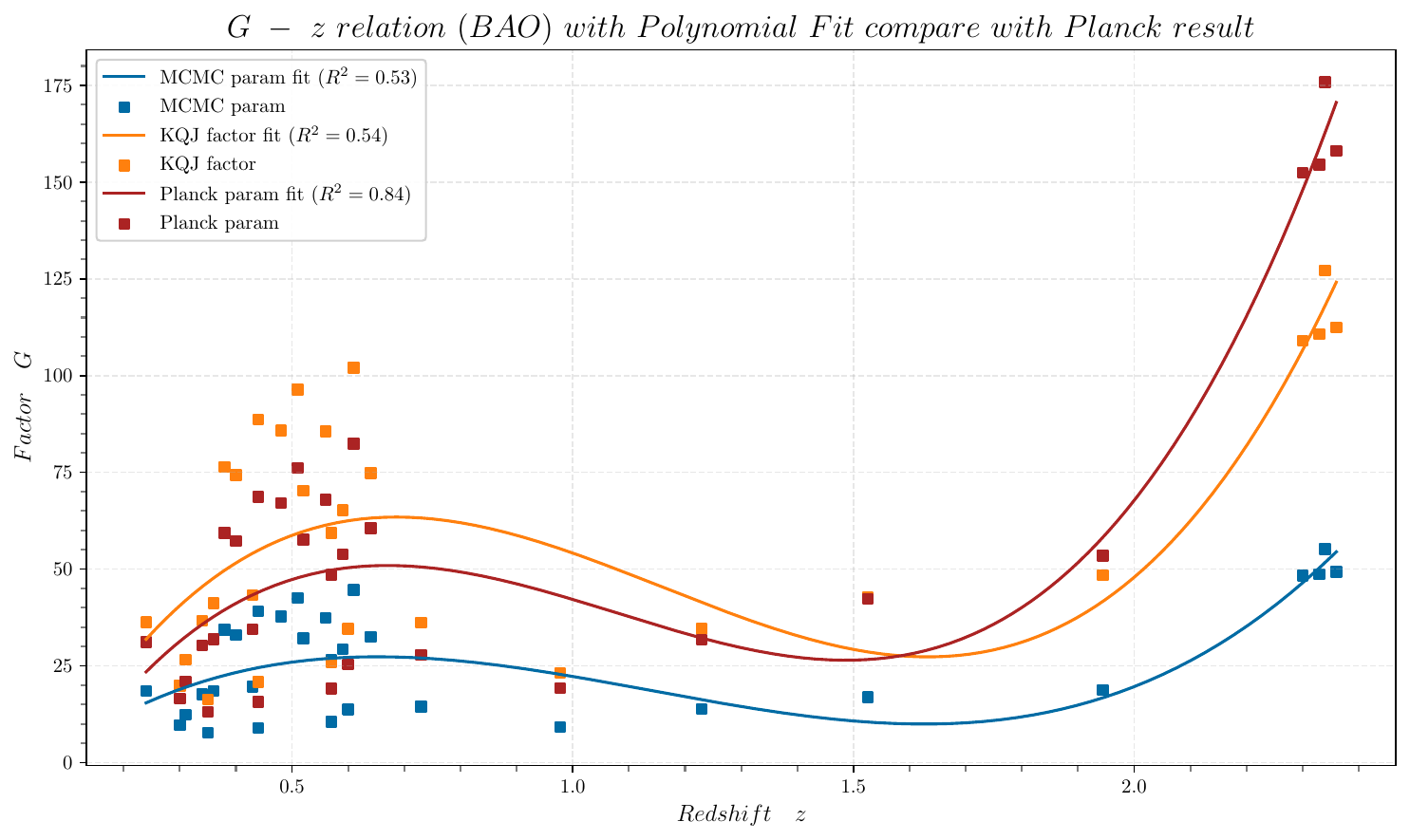}
   \caption{illustrates the cubic fitting curves of the \( G \) factor derived using three different methods for cosmological parameter estimation. The blue curve corresponds to parameters obtained from MCMC analysis, the green curve represents results directly provided in \cite{Planck2020}, and the red curve is constructed based on three parameters defined in the text, which are related to \( G \) and its higher-order derivatives with respect to \( z \).}
   \label{Fig4}
   \end{figure*}

To provide a theoretical explanation for the cubic curve shape of the $G$ factor derived from the BAO dataset (while the $G$ factor for the CC data exhibits a near-linear distribution, more consistent with physical intuition and thus not discussed further), we draw inspiration from the definition of the deceleration parameter $q$ in cosmology and extend this concept. The following factors related to $G$ are defined:
\begin{equation}
    K(z) = \frac{\dot{G}(z)}{G(z)}, \quad Q(z) = \frac{\ddot{G}(z)G(z)}{\dot{G}^{2}(z)}, \quad J(z) = \frac{\dddot{G}(z)G^{2}(z)}{\dot{G}^{3}(z)}.
\end{equation}
Here, the relationship between \( K(z) \) and \(G\) is analogous to the Hubble parameter \(H(z)\) and the scale factor \(a\), \( Q(z) \) is similar to the deceleration factor \( q \), and \( J(z) \) represents higher-order deceleration factors. With the above definitions, the \( G \) factor can be expressed in the following expanded form at \(z_{0}\):
\begin{align}
    G(z) = G(z_{0})\{ 1+K(z_{0})z + \frac{1}{2!} Q(z_{0})K^{2}(z_{0})z^2 + \notag
    \\ \frac{1}{3!}J(z_{0})K^{3}(z_{0})z^3 + O(z^4) \}.
    \label{G_expand}
\end{align}
Fig.~\ref{Fig4} shows the results of polynomial fitting performed using the \( G \) factor calculated from the above expansion \eqref{G_expand} at various data points. These results are compared with the polynomial fitting curves obtained by using cosmological parameters derived from the $\Lambda$CDM model, either input from Planck result(green) or obtained via MCMC simulations using the BAO dataset(blue).

A notable observation is that the regression coefficient $R$ of the fitting curve obtained using the expanded form of $G$ closely approximates that derived from the BAO dataset. This suggests that, under the defined conditions, the information content of the BAO dataset is effectively encapsulated by the three factors $K(z)$, $Q(z)$, and $J(z)$. Furthermore, the resulting fitting curve exhibits a reasonable agreement in $G$ factor values with that obtained using the Planck parameter set. However, a discrepancy remains in the regression coefficients, potentially attributable to errors introduced by neglecting higher-order terms, which may distort the fitting process.

\subsection{Effectiveness analysis for $G$}
The above discussion is based on the premise that the \( G \) factor is a suitable evaluation quantity. The following section presents an effectiveness analysis to demonstrate that the \( G \) factor indeed serves to distinguish the datasets.

\begin{table*}[ht!]
\centering
\caption{employs three different datasets (CC, BAO, and CC+BAO), each further divided into two subsets based on the magnitude of the \( G \) factor (high and low), resulting in a total of \( 2 \times 3 = 6 \) parameter sets. The table lists the values of each element in the FoM matrix calculated sequentially according to Eq.\eqref{FoM_matrix_element}, as well as the absolute value of the determinant of the matrix.
We emphasize here again that all the following calculations are all based on the $\Lambda$CDM model.}
\label{FoM data}
\begin{tabular}{ccccccc}
\hline
\textbf{Dataset} & \textbf{CC low $G$ } & \textbf{CC high $G$} & \textbf{BAO low $G$} & \textbf{BAO high $G$} & \textbf{CC+BAO low $G$} & \textbf{CC+BAO high $G$}\\ 
\hline
Index & a & b & c & d & e & f \\
\hline
$FoM_{H\Lambda}$ & $1.87 \times 10^0$ & $2.07 \times 10^1$ & $2.14 \times 10^1$ & $1.00 \times 10^2$ & $7.99 \times 10^0$ & $1.27 \times 10^2$ \\
$FoM_{Hm}$ & $8.55 \times 10^{-1}$ & $7.15 \times 10^0$ & $9.99 \times 10^0$ & $6.22 \times 10^1$ & $4.01 \times 10^0$ & $8.01 \times 10^1$ \\
$FoM_{Hk}$ & $2.03 \times 10^0$ & $6.38 \times 10^0$ & $1.92 \times 10^1$ & $8.20 \times 10^1$ & $7.55 \times 10^0$ & $1.07 \times 10^2$ \\
$FoM_{\Lambda m}$ & $3.05 \times 10^1$ & $3.86 \times 10^2$ & $4.96 \times 10^2$ & $3.54 \times 10^3$ & $1.79 \times 10^2$ & $4.45 \times 10^3$ \\
$FoM_{\Lambda k}$ & $4.76 \times 10^1$ & $4.36 \times 10^2$ & $6.54 \times 10^2$ & $3.06 \times 10^3$ & $2.19 \times 10^2$ & $3.76 \times 10^3$\\
$FoM_{mk}$ & $4.99 \times 10^1$ & $4.39 \times 10^2$ & $8.25 \times 10^2$ & $4.81 \times 10^3$ & $2.62 \times 10^2$ & $5.84 \times 10^3$ \\
$|det(FoM)|$ & $9.97 \times 10^3$ & $1.85 \times 10^7$ & $2.46 \times 10^8$ & $2.21 \times 10^{11}$ & $4.72 \times 10^6$ & $5.76 \times 10^{11}$\\
\hline
\end{tabular}
\end{table*}

\begin{table*}[ht!]
\centering
\caption{presents a unified comparison of cosmological parameter constraints derived from SNIa, Planck, and different $G$ subsets of the combined CC+BAO dataset. The Hubble constant $H_{0}$ is expressed in units of $\mathrm{km\ s^{-1}\ Mpc^{-1}}$.}
\label{compare}
\begin{tabular}{ccccc}
\hline
\textbf{Dataset} & \textbf{$H_{0}$ } & \textbf{$\Omega_{\Lambda}$} & \textbf{$\Omega_{m}$} & \textbf{$\Omega_{k}$} \\ 
\hline
SNIa--Pantheon+ \cite{Pantheon+} & $73.4^{+1.1}_{-1.1}$ & $0.625^{+0.084}_{-0.084}$ & $0.306^{+0.057}_{-0.057}$ & / \\
Planck \cite{Planck2020} & $67.36^{+0.54}_{-0.54}$ & $0.6847^{+0.0073}_{-0.0073}$ & $0.3153^{+0.0073}_{-0.0073}$ & $-0.011^{+0.013}_{-0.012}$ \\
CC+BAO \cite{wu2024cosmicdynamicseinsteincartantheory} & $67.6_{-2.2}^{+2.7}$ & $0.681_{-0.047}^{+0.037}$ & $0.320_{-0.038}^{+0.046}$ & $0.001_{-0.054}^{+0.053}$ \\
CC+BAO (Low $G$) & $68.9^{+2.9}_{-3.1}$ & $0.724^{+0.063}_{-0.074}$ & $0.283^{+0.065}_{-0.058}$ & $-0.002^{+0.053}_{-0.055}$ \\
CC+BAO (High $G$)& $70.31^{+0.73}_{-0.83}$ & $0.738^{+0.017}_{-0.014}$ & $0.259^{+0.012}_{-0.011}$ & $0.003^{+0.012}_{-0.014}$ \\
\hline
\end{tabular}
\end{table*}

To validate the effectiveness of the $G$ factor in mitigating model degeneracies within the dataset, we employ a data partitioning strategy based on $G$ factor magnitude. Data points are categorized into low $G$ and high $G$ subsets, with the data point possessing the lowest $G$ factor removed if the dataset contains an odd number of points. Subsequently, MCMC analysis is performed on each subset, and the resulting parameters and their associated uncertainties $\sigma$ are used to construct the FoM matrix. The magnitude of each element within the FoM matrix serves as a metric for dataset quality. Table.~\ref{FoM data} presents the FoM values computed from the $G$ factor-based partitioning of three distinct datasets: CC, BAO, and CC+BAO. As shown, with the exception of a few outliers, the FoM values for any parameter combination within the high $G$ group consistently exceed those in the low $G$ group (i.e., $b > a$, $d > c$, and $f > e$), strongly supporting the validity of the $G$ factor as a quality indicator.

We now turn our attention to the final row of Table.~\ref{FoM data}, which displays the determinant of the FoM matrix. It is generally accepted that this value reflects the volume of the accessible parameter space; a larger determinant indicates a higher quality parameter set. Therefore, it is included as a key indicator of degeneracy. Examination of the table reveals that the $det(\text{FoM})$ for the different datasets follows the order: $c > e > a$ and $f > d > b$. These sequences underscore a crucial point: simply increasing the number of data points used in the fitting does not guarantee a reduction in fitting degeneracies. In some cases, incorporating low-quality data can even have a detrimental effect. Consequently, prioritizing datasets with higher $G$ values is essential when performing fitting calculations.

   \begin{figure*}[ht!]
   \centering
   \includegraphics[width=0.85\textwidth, angle=0]{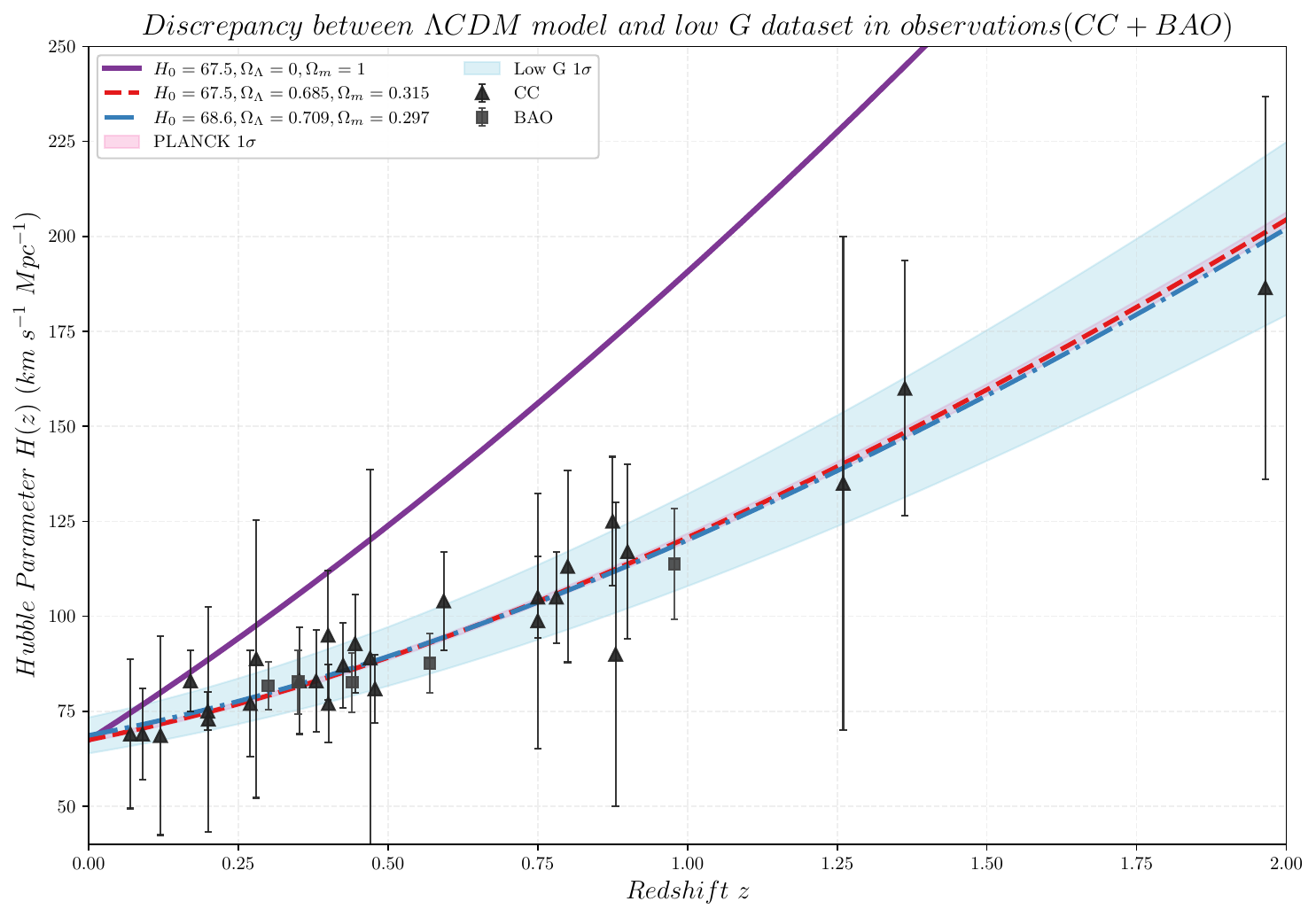}
   \caption{The three curves in Fig.~\ref{Fig5} illustrate the evolution of the Hubble parameter \( H \) with redshift \( z \) under different cosmological parameter settings. The red curve represents a purely baryonic universe, the blue curve corresponds to the cosmological model based on Planck parameters, and the green curve reflects the universe model derived from the MCMC analysis of the low \( G \) parameter set. The shaded regions indicate the corresponding 1$\sigma$ uncertainty intervals.}
   \label{Fig5}
   \end{figure*}

   \begin{figure*}[ht!]
   \centering
   \includegraphics[width=0.85\textwidth, angle=0]{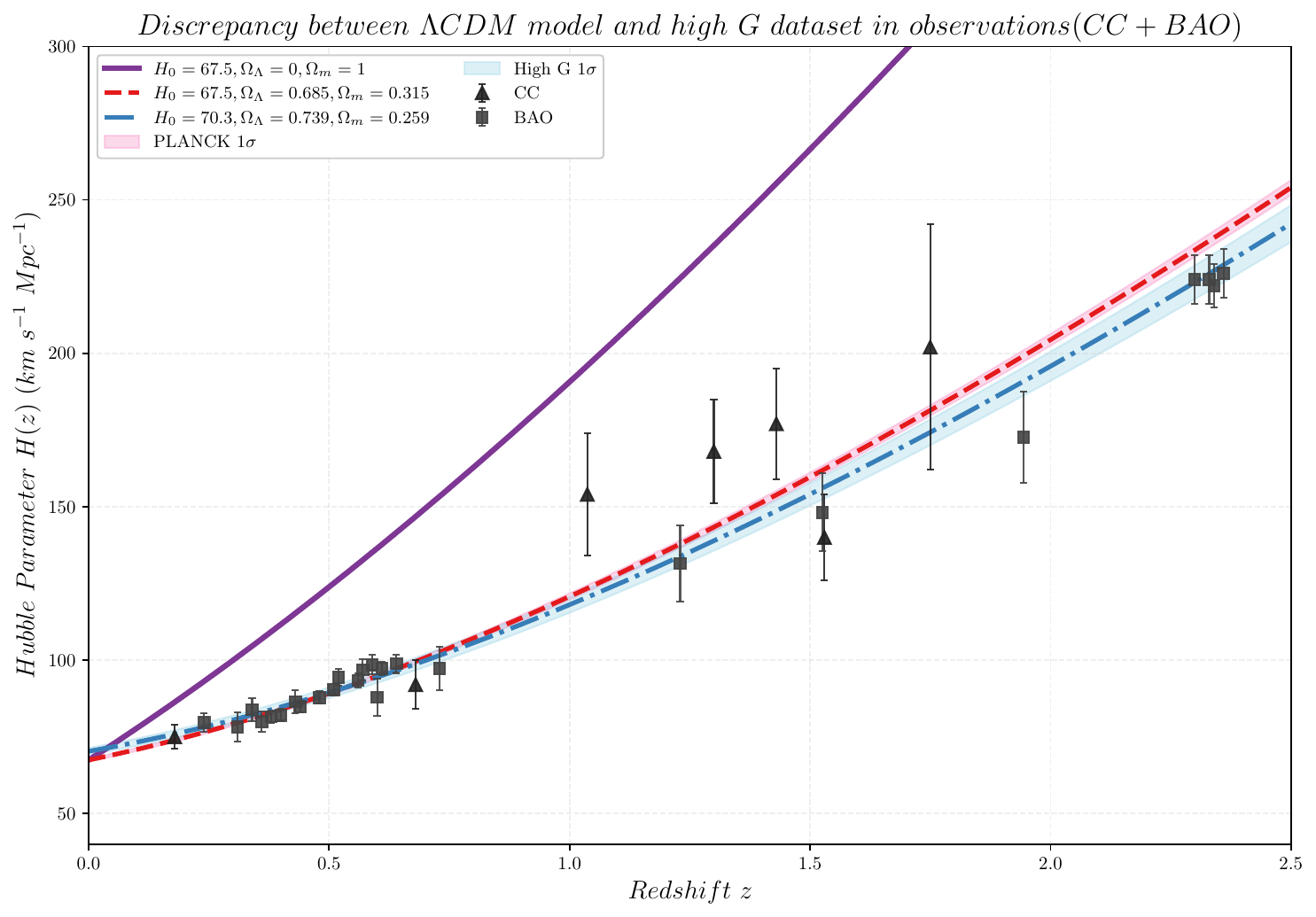}
   \caption{The plotting principles of Fig.~\ref{Fig6} are the same as those of Fig.~\ref{Fig5}, but the dataset used for the MCMC analysis is based on the high \( G \) subset.}
   \label{Fig6}
   \end{figure*}

To further elaborate, we compare the Hubble parameter \( H \) curves derived from fitting the two $G$ factor partitions of the CC+BAO dataset, as illustrated in Fig.~\ref{Fig5} and Fig.~\ref{Fig6}. 
The purple curve represents a purely baryonic universe serving as a null control group and the red curve is obtained by direct fitting using the cosmological parameters from Planck, while the blue curve is derived from the MCMC fitting results based on the different $G$ factor partitions. The semi-transparent regions correspond to the $1\sigma$ confidence intervals of the respective \( H \) values, calculated using the error propagation formula.

It can be observed that the light green region in Fig.~\ref{Fig5} is significantly larger than both the light red region in Fig.~\ref{Fig5} and the light blue region in Fig.~\ref{Fig6}. Within the $1\sigma$ error range of the data points in Fig.~\ref{Fig5}, it is difficult to identify the parameter trends. In contrast, in Fig.~\ref{Fig6}, the coverage areas of the red and blue regions are more similar, with distinct non-overlapping sections. This visually demonstrates that the high $G$ dataset provides better fitting quality and higher parameter discriminability.

Finally, to gain an intuitive understanding of the screening efficacy of $G$, we present a comparative analysis of the cosmological parameter constraints derived from the SNIa, Planck, and CC+BAO datasets (separated into High $G$ and Low $G$ subsets) in Table.~\ref{compare}. It is readily apparent that the constraint precision of the High $G$ group closely approaches that of Planck, whereas the Low $G$ group exhibits significantly reduced precision, falling considerably below the constraints from SNIa. This demonstrates that $G$ serves as an effective screening criterion. 

Readers may note a tension between the Hubble parameter $H_{0}$ and density parameters of the first two and the last two rows containing the $G$ factor groups in Table.~\ref{compare}, compared to the parameters within the first two rows themselves, which is a phenomenon worth for explanation. It can be observed that the parameters of the third row, which contains no $G$ factor, are very close to the Planck data. As the magnitude of the $G$ factor increases, the discrepancy with the Planck data also increases. This is essentially due to the dependence of the $G$ factor itself on data precision. Current observational methods yield higher precision at low redshifts, where smaller relative measurement errors correspond to larger $G$ factors. Consequently, the high $G$ subgroup naturally exhibits a shift toward the cosmological parameters derived from SNIa constraints, compared to the third row with zero $G$ factor as a control. In contrast, the deviation in the low $G$ subgroup is not caused by the aforementioned reason but is more likely a result of increased final constraint errors due to the removal of high-precision measurement data, leading to a certain degree of dispersion in the central values.

This suggests that the $G$ factor does not have inherent physics related to the Hubble tension and is primarily applicable for decrease discrepancy between models and data screening.

\subsection{Results from simulated OHD dataset}
\begin{figure*}[ht!]
    \centering
    \begin{tabular}{cc}
        \includegraphics[width=0.43\textwidth]{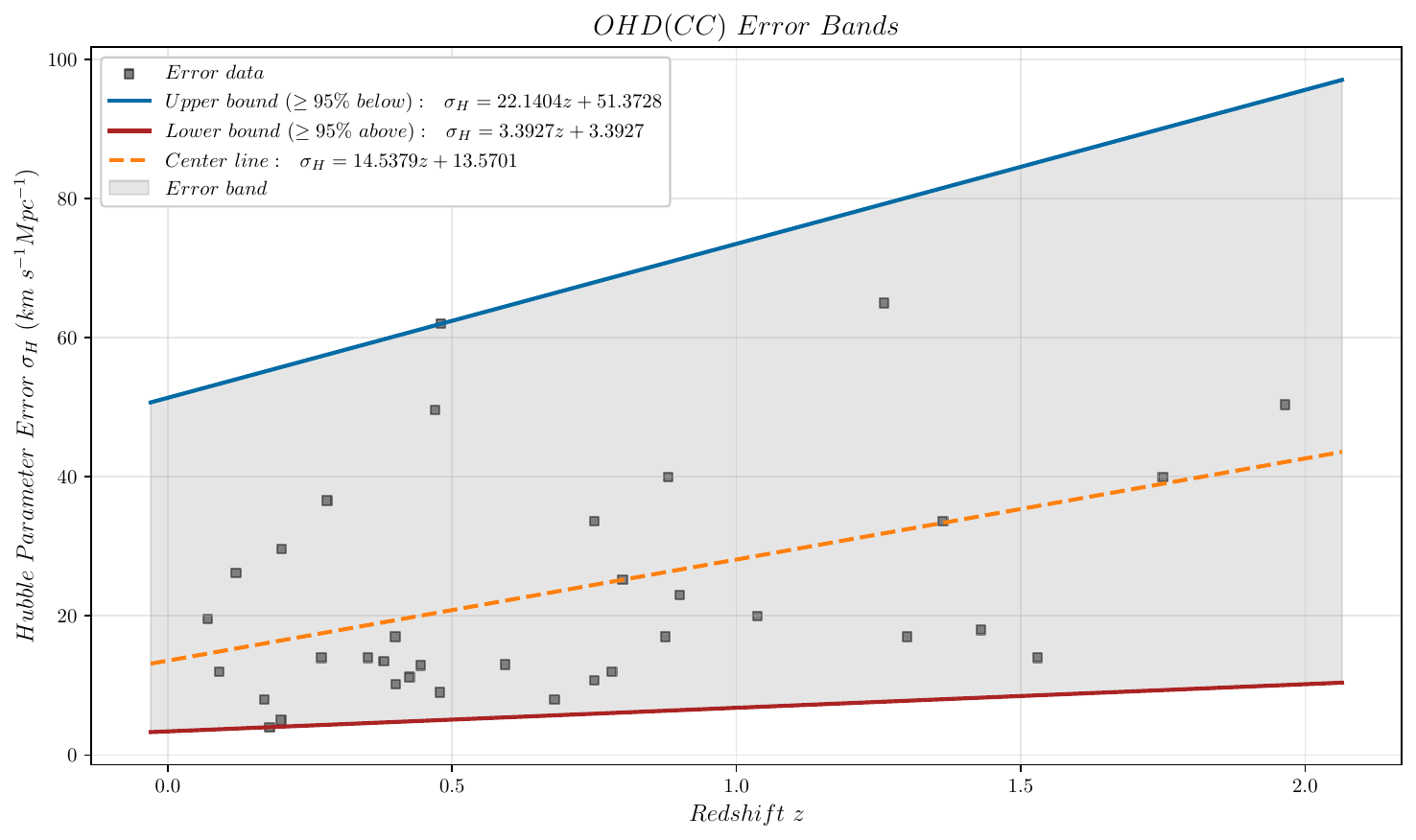} &
        \includegraphics[width=0.43\textwidth]{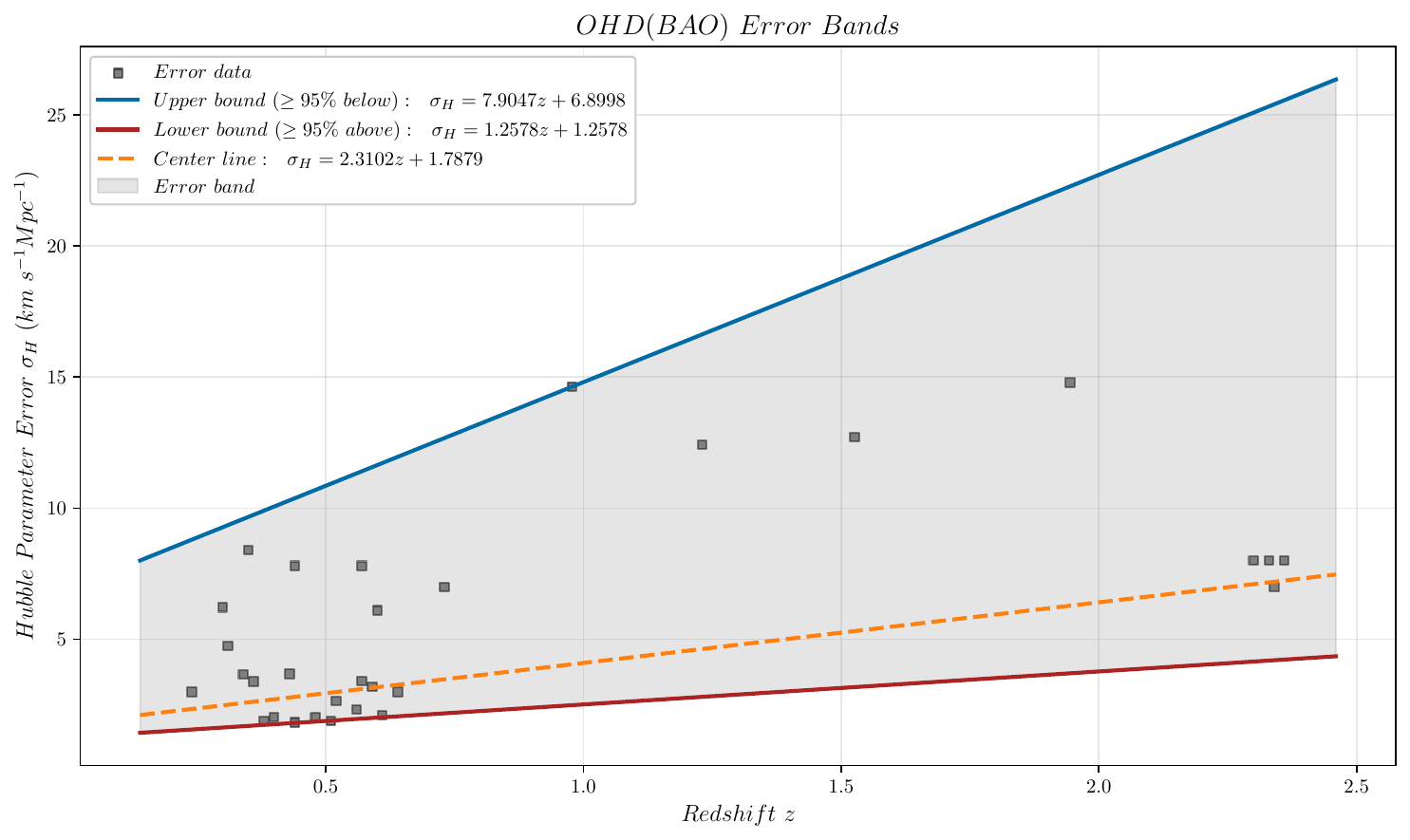}
    \end{tabular}
    \caption{plots the error distribution of the Hubble parameter measurements derived from both the CC and BAO datasets. Using the rejection sampling method, linear upper and lower bounds were determined. The grey enveloping region represents the possible range of errors, while the central line depicts a linear regression fitted to the measured error points within the figure.}
    \label{error band}
\end{figure*}

   \begin{figure*}[ht!]
   \centering
   \includegraphics[width=0.85\textwidth, angle=0]{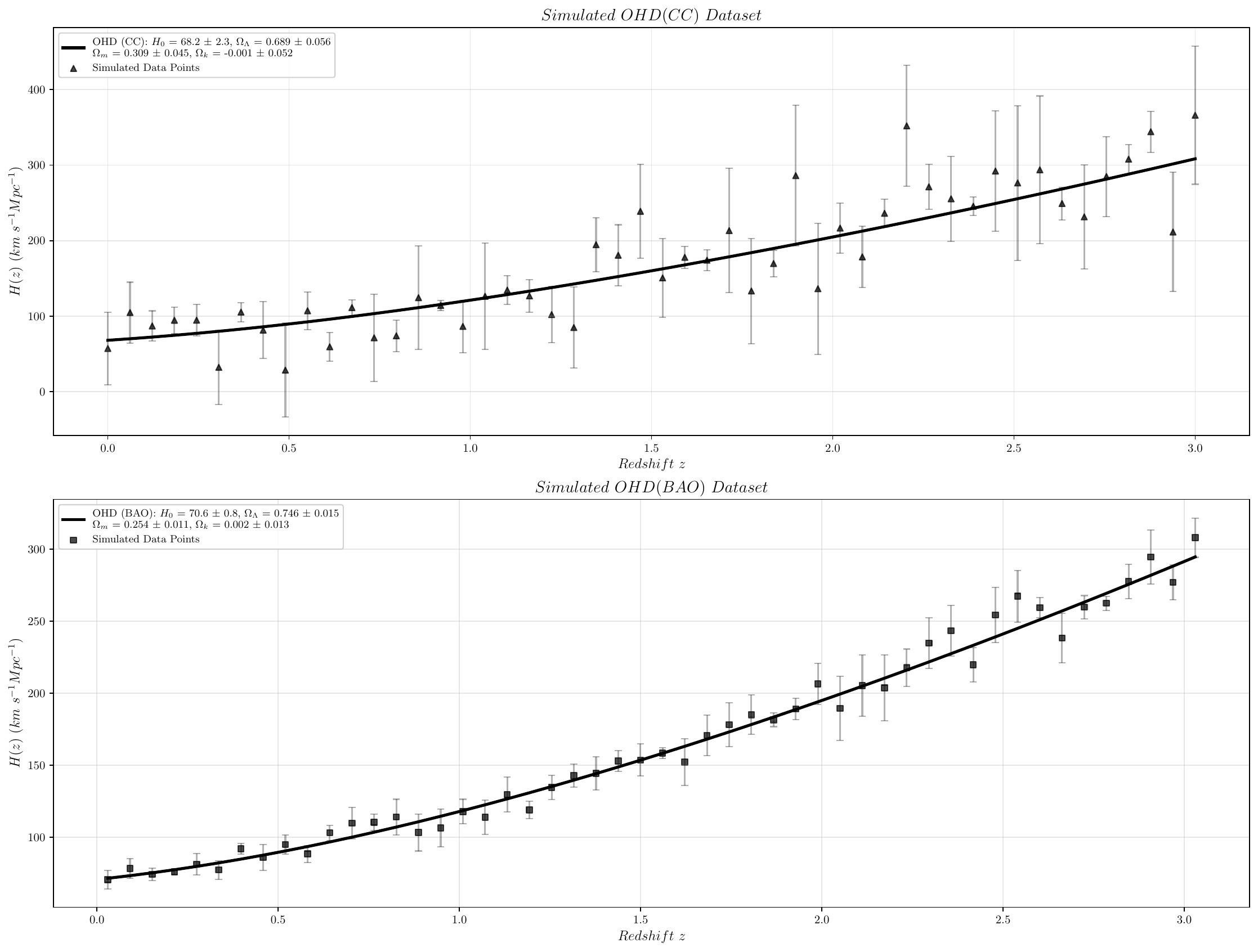}
   \caption{presents simulated OHD generated using the method outlined for Fig.~\ref{error band}. The simulations are categorized and shown separately as CC-like and BAO-like types, due to their significantly different predicted characteristics.}
   \label{OHD data}
   \end{figure*}

   \begin{figure*}[ht!]
   \centering
   \includegraphics[width=0.85\textwidth, angle=0]{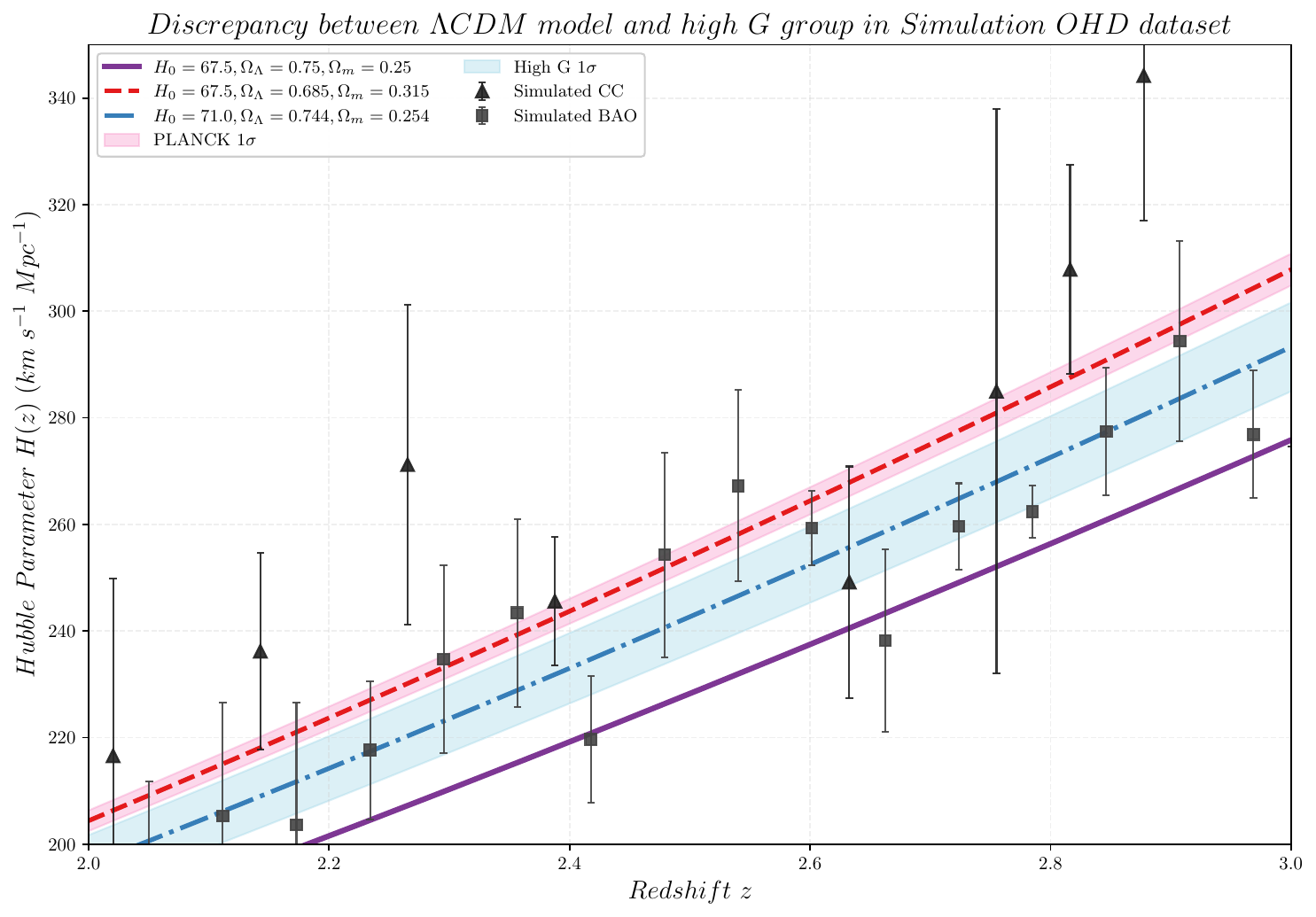}
   \caption{employs a construction logic analogous to Fig.\ref{Fig6}. However, it utilizes the simulated OHD data in this instance. Furthermore, since the best-fit curve corresponding to a pure baryonic universe exhibits a marked deviation at high redshifts, it has been replaced by the purple reference curve. This curve, generated using cosmological parameters for the $\Lambda CDM$ model constrained by SNIa data, provides a more meaningful comparison.}
   \label{OHD discrepancy}
   \end{figure*}
   
To ensure the $G$ factor can effectively break model degeneracy in future observations, we validate its performance using simulated OHD datasets (where OHD corresponds to the CC + radial BAO compilation), following the fitting methodology detailed in \cite{Ma_2011}. Fig.~\ref{error band} presents the predicted uncertainty ranges derived from OHD data, while Fig.~\ref{OHD data} displays the corresponding simulated OHD data. In Fig.~\ref{OHD discrepancy}, a clear distinction emerges between the Planck parameters and those derived from simulated OHD data (with 1:1 input ratio of simulated CC to simulated BAO data, 100 points in total) at $z > 2.5$. This divergence extends the separation between fitting methods observed in Fig.~\ref{Fig6}, which exhibiting approximately linear growth with redshift. These results robustly demonstrate the capacity of high $G$ data to resolve degeneracy in high redshift regimes.
Furthermore, to test whether degeneracy issues might be mitigated through reduced measurement errors in future OHD surveys, we repeat the MCMC analysis using simulated OHD data with halved observational uncertainties. The constraints results are shown below:
\\Original Simulated OHD:
\\$H_{0}= 71.1^{+1.2}_{-1.2}$, $\Omega_{\Lambda}=0.744^{+0.018}_{-0.017}$, $\Omega_{m}=0.253^{+0.011}_{-0.011}$, $\Omega_{k}=0.002^{+0.017}_{-0.017}$;
\\Reduced uncertainty Simulated OHD:
\\$H_{0}= 71.14^{+0.62}_{-0.61}$, $\Omega_{\Lambda}=0.7454^{+0.0084}_{-0.0088}$, $\Omega_{m}=0.2525^{+0.0084}_{-0.0088}$, $\Omega_{k}=0.0009^{+0.0095}_{-0.0074}$;
\\It is readily observed that the ratio between the uncertainty of the parameter constraints and the error of the simulated OHD data is essentially identical to the reduction ratio of the observational errors. Meanwhile, the central values of the cosmological parameters show negligible variation. This demonstrates the viability of resolving model degeneracies through improved precision in OHD measurements.

\section{Conclusions}

In summary, this study has reached the following key findings:  

1. The $G$ factor of the CC dataset generally increases linearly with redshift \( z \), with the linear regression coefficient \( R^2 = 0.56 \) for its dominant component, indicating a moderate correlation. Therefore, future additions to the CC dataset should ideally focus on higher redshifts to determine whether the linear model remains valid.

2. The $G$ factor of the BAO dataset exhibits a cubic relationship with redshift \( z \), first increasing, then decreasing, and finally increasing again. The linear regression coefficient \( R^2 = 0.90 \) for its dominant component indicates a strong correlation, suggesting that the fitted coefficient holds physical significance. Future work could focus on identifying the theoretical counterparts of the three factors, $K$, $Q$, and $J$, in the BAO formation theory. Additionally, for future observations aiming to minimize redshift for practical reasons, it is advisable to target objects near the first extremum of this cubic relationship, with a redshift \( z \approx 0.6 \).

3. Under the evaluation system using the $G$ factor as a criterion, the BAO dataset yields better parameter constraints than the CC dataset. Quantifying the comparison with the FoM value, we find that for the $G$ factor, the ratio \( \text{FoM}_{BAO}/\text{FoM}_{CC} \sim 10^{1-2} \).

4. In the framework where the FoM is used as an evaluation criterion, the $G$ factor effectively serves to filter high quality observational points within the dataset. For CC+BAO this combined dataset, the curve $1\sigma$ area plotted from the high $G$ group after selection is approximately $\frac{1}{10}$ as broad as that of the low $G$ group. As for the specific cosmological parameters, statistically, datasets with a higher $G$ factor tend to produce parameter constraints that are closer to the results provided by \cite{Planck2020}.

5. Crucially, this study advances beyond the work of \cite{2016RAA....16...50T} by not only justifying the development of the $G$ factor into an observational diagnostic using the $BIC$, but also validating its utility through simulated OHD. Specifically, we demonstrate that the $G$ factor more effectively breaks the degeneracy among cosmological models at higher redshifts ($z > 2.5$). We anticipate that future work will extend the application of the $G$ factor beyond OHD datasets to large-scale surveys featuring objects with high observational precision—such as DESI and LSST thereby establishing it as a more universally applicable classification criterion.

\section*{Acknowledgements}
This work was supported by National SKA Program of China, No.2022SKA0110202 and the China Manned Space Program with grant No.CMS-CSST-2025-A01.

\appendix

\section{The mathematical form of Figure of Merit}
This section will derive the specific mathematical form of the FoM starting from the definition upon. According to statistical conclusions, for two parameters $ \{X,Y\}$ obey Gaussian distributions and correlation coefficient $\rho \equiv \rho_{XY}$, their joint probability density function $f_{X,Y}$ can be written as:
\begin{align}
    f_{X,Y}(x,y) = \frac{1}{2\pi\sigma_{X}\sigma_{Y}\sqrt{1-\rho^{2}}}\exp(-\frac{1}{2(1-\rho^{2})} \notag
    \\ [\frac{(x-\mu_{X})^2}{\sigma_{X}^{2}} + \frac{(y-\mu_{Y})^2}{\sigma_{Y}^{2}} - \frac{2\rho(x-\mu_{X})(y-\mu_{Y})}{\sigma_{X}\sigma_{Y}} ]),
\end{align}
where the $\mu_{X},\mu_{Y},\sigma_{X},\sigma_{Y}$ are the mean and variance of $\{X,Y\}$, respectively. By observing the index part of the exponential term, the covariance matrix $\Sigma$ of this parameter combination can be directly written as:
\begin{equation}
    \Sigma = 
          \begin{pmatrix}
                \sigma_{X}^{2} & \rho\sigma_{X}\sigma_{Y} \\
                \rho\sigma_{X}\sigma_{Y} & \sigma_{Y}^{2}
          \end{pmatrix}.
\end{equation}

Returning to the definition, the 95\% confidence interval constraint actually creates a contour on the surface of the joint Gaussian distribution. The region enclosed by this contour is the ellipse in the definition. The directions of the major and minor axes $a,b$ of this ellipse should be parallel to the eigenvalue vectors of the covariance matrix. Therefore, the problem becomes to solve the following equation:
\begin{align}
    det(\Sigma-\lambda I) = 
    (\sigma_{X}^{2}-\lambda)(\sigma_{Y}^{2}-\lambda)-\rho^{2}\sigma_{X}^{2}\sigma_{Y}^{2}\notag
    \\ = \lambda^{2} - (\sigma_{X}^{2} + \sigma_{Y}^{2})\lambda + (1-\rho^{2})\sigma_{X}^{2}\sigma_{Y}^{2} = 0.
\end{align}

After the above simplification, according to Vieta theorem, the product of the two different eigenvectors $\vec{\lambda_{1}},\vec{\lambda_{2}}$ satisfies:
\begin{equation}
    \lambda_{1}\lambda_{2} = (1-\rho^{2})\sigma_{X}^{2}\sigma_{Y}^{2} \propto a^{2}b^{2}.
\end{equation}

Therefore, the area $S$ of the ellipse in the definition is proportional to the following equation:
\begin{equation}
    S = \pi ab \propto \pi\sqrt{(1-\rho^{2})}\sigma_{X}\sigma_{Y} = A,
\end{equation}
where $A$ is the area of the ellipse formed by the two eigenvectors. Finally, in order to maintain the simplicity of the mathematical expression (since FoM is a statistic used for comparison, absolute size of it does not affect its physical meaning), we make the following definition:
\begin{equation}
    FoM_{XY} = \frac{\pi}{A} = \frac{1}{\sqrt{(1-\rho^{2})}\sigma_{X}\sigma_{Y}}.
\end{equation}

\bibliographystyle{elsarticle-harv} 
\bibliography{citations}

\end{document}